 \documentclass[final,1p,times,twocolumn]{elsarticle}

\usepackage{amssymb}
\usepackage{amstext}
\usepackage{amsmath}
\usepackage[table]{xcolor}
\usepackage{color}
\usepackage{lscape}
\usepackage{multirow}
\usepackage{arydshln}
\usepackage{url,floatflt}
\usepackage{subfig}
\usepackage{rotating}
\usepackage{amsfonts}
\usepackage{cancel}
\usepackage{amssymb}
\usepackage{stackrel}
\usepackage{tikz}
\usepackage{slashbox}
\usetikzlibrary{shapes,snakes}
\newcommand*\circled[1]{\tikz[baseline=(char.base)]{
            \node[shape=circle,draw,inner sep=2pt] (char) {#1};}}
\newcommand*\circledr[1]{\tikz[baseline=(char.base)]{
            \node[shape=circle,draw,inner sep=2pt,fill=red!30] (char) {#1};}}
\newcommand*\circledo[1]{\tikz[baseline=(char.base)]{
            \node[shape=circle,draw,inner sep=2pt,fill=yellow!30] (char) {#1};}}
            \newcommand*\circledb[1]{\tikz[baseline=(char.base)]{
            \node[shape=circle,draw,inner sep=2pt,fill=blue!30] (char) {#1};}}
            \newcommand*\circledg[1]{\tikz[baseline=(char.base)]{
            \node[shape=circle,draw,inner sep=2pt,fill=green!30] (char) {#1};}}
            \newcommand*\circledbl[1]{\tikz[baseline=(char.base)]{
            \node[shape=circle,draw,inner sep=2pt,fill=black!30] (char) {#1};}}

\newcommand{\Fset}{\mathbb{F}}
\newcommand{\zech}[2]{\mathcal{Z}_{#1}\left({#2}\right)}

\usepackage[amsmath,thmmarks]{ntheorem}
\theoremheaderfont{\bf}
\theorembodyfont{\itshape}
\theoremstyle{plain}
\theoremseparator{:}
\theoremsymbol{}

\newtheorem{theorem}{Theorem}
\newtheorem{corollary}{Corollary}

\newtheorem{remark}{Remark}

\theorembodyfont{\upshape}
\newtheorem{definition}{Definition}
\theoremsymbol{\ensuremath{\Box}}
\newtheorem{example}{Example}

\theoremheaderfont{\sc}
\theorembodyfont{\upshape}
\theoremstyle{nonumberplain}
\theoremsymbol{\ensuremath{\square}}
\newtheorem{proof}{Proof}
\usepackage{xfrac}
 \usepackage[symbol]{footmisc}

\journal{arXiv}

\begin{document}

\begin{frontmatter}



\title{Cellular Automata as Generators of Interleaving Sequences}

 \author[label1]{S. D. Cardell \footnote{Corresponding author}}
 \ead{sd.cardell@unesp.br}
 \address[label1]{São Paulo State University (Unesp)\\ Institute of Geosciences and Exact Sciences\\ Rio Claro, Brazil}

\begin{abstract} 
An interleaving sequence is obtained by  combining or intertwining elements from two or more sequences. 
On the other hand, cellular automata are known to be generators for keystream sequences. 
In this paper we present two families of   one-dimensional cellular automata as generators of interleaving sequences. 
This study aims to close a notable gap within the current body of literature by exploring the capacity of cellular automata to generate interleaving sequences. While previous works have separately examined cellular automata as sequence generators and interleaving sequences, there exists limited literature interconnecting these two topics. Our study seeks to bridge this gap, providing perspectives on the generation of interleaving sequences through the utilisation of cellular automata, thereby fostering a deeper understanding of both disciplines.

\end{abstract}


\end{frontmatter}


\section{Introduction}\label{sec:intro}
Binary sequences generated by maximal-period Linear Feedback Shift Registers (LFSRs), known as PN-sequences  \cite{Golomb1982bk}, find extensive applications in various fields like digital broadcasting, mobile wireless communications and cryptography (specifically in stream ciphers). To enhance practical cryptographic stability, it becomes essential to eliminate the inherent linearity of PN-sequences by incorporating nonlinear procedures.


Efficient sequence generation is achieved through the use of Linear Feedback Shift Registers (LFSRs), making them well-suited for various cryptographic applications, for example,
they were used in algorithms such us
the A$5$  for GSM communications \cite{Cardell2019bk}, the RC4 algorithm employed in encrypting Internet traffic \cite{Paul2012bk}, the Grain-128AEAD candidate in the NIST Lightweight Crypto Standardization process \cite{NIST2019} or the Trivium \cite{Canniere2006}.
Other recent algorithms that used LFSRs in their design are Lizard and Flip \cite{Hamann2017,Duval2016}.
Among the prominent families of cryptographic sequence generators, irregular decimation-based generators stand out \cite{Cardell2019bk}.
 
The method used in the irregularly decimating of the output sequence of an LFSR is a powerful tool to construct sequences with good cryptographic properties such as: long  periods, good distribution of zeros and ones along the sequence, large linear complexity, good auto-correlation properties, etc. 
Among all irregularly decimated generators, we can highlight the shrinking generator \cite{Coppersmith1993} composed of two LFSRs with different lengths, where one PN-sequence decimates the other. 
This generator is fast, easy to be implemented and generates good cryptographic sequences, so they seem adequate for their use in light-weight cryptography and, in general, in low-cost applications.
In \cite{Cardell2016f}, the authors showed that the sequences obtained with the shrinking generator can be also produced interleaving PN-sequences. In fact, these PN-sequences are all generated by the same primitive polynomial, that is, they are shifted versions of the same PN-sequence. 

In \cite{Cardell2020c}, the authors computed
the shifts of the interleaving PN-sequences in the shrunken sequence. 
This fact can be used advantageously to perform  attacks against the shrinking generator \cite{Cardell2020c}.
A natural way to deal with this liability is to randomise the shifts.
In  \cite{Cardell2021a} the authors studied the resultant sequences of
interleaving shifted versions of the same PN-sequence with different shifts. They analyse the
conditions these shifts must satisfy   to obtain interleaving sequences with good cryptographic potential, such as high linear
complexity and long period.
 

On the other hand, cellular automata  are known for producing keystream sequences \cite{Cardell2016f,Wolfram1986,SARKAR2003,Cardell2016c,Cardell2016d,Cardell2017b}. Notably, the works of \cite{Cardell2016f, Fuster2007c} demonstrated that the output sequences from the shrinking generator can be derived as vertical sequences from different families of cellular automata. It is reasonable to conjecture that there exist families of cellular automata   responsible for generating the resulting sequences of interleaving random shifted versions of a PN-sequence. In this study, we examine two categories of linear cellular automata that generate interleaving sequences, obtained interlacing shifted versions of the same PN-sequence.

This paper is organised as follows.
In Section~\ref{sec:prel} we introduce some basic notation and several notions needed to understand the rest of the paper.
In Section~\ref{sec:mismo}, we present the main results. 
We study the structure of the CAs that generate interleaving sequences obtained weaving shifted versions of the same PN-sequence.
In Section~\ref{sec:comp}, we discuss the different characteristics of the various families of cellular automata that generate the interleaving sequences.
Finally, the paper concludes in Section~\ref{sec:concl} with some conclusions and future work.

\section{Fundamentals and basic notation}\label{sec:prel}
 In this section, we introduce fundamental concepts essential for a complete understanding of our work. 
\subsection{Characteristics of the binary sequences}\label{sec:binary}
Let $\mathbb{F}_2$ be the Galois field of two elements (or binary field).
The sequence $\{a_i\}_{i\geq 0}=\{a_0, a_1, a_2 \ldots\}$ is a binary sequence if $a_i\in \mathbb{F}_2$,
for $i=0, 1, 2,\ldots$. 
From now on, all   sequences considered in this work will be binary sequences and  $+$ will denote the Exclusive-OR (XOR) logic operation and the multiplication $\cdot$ will be AND logic operation.  
The sequence $\{a_i\}_{i\geq 0}$ is periodic if and only if there exists an integer $T$ such that    $a_{i+T}=a_i$ for all $i\geq 0$.

Let $p_0, p_1, p_2, \ldots, p_{L-1}\in\mathbb{F}_2$ be constant coefficients and $L$  a positive integer.
A binary sequence $\{a_i\}_{i\geq 0}$ (or simply $\{a_i\}$) that satisfies the relation
\begin{equation}\label{eq:1}	
a_{i+L}=p_{L-1}a_{i+L-1} + \cdots +p_{2}a_{i+2}+p_{1}a_{i+1}+p_0a_{i}, \quad i\geq 0,
\end{equation}
is an $L$-th order linear recurring sequence defined over $\mathbb{F}_2$.
The first  $L$ terms of the sequence $\{a_0, a_1, \ldots,$ $a_{L-1}\}$ are called the initial state and uniquely determine the remaining bits of the sequence.
The equation~(\ref{eq:1}) denotes an $L$-th order linear recurrence relationship.

The monic polynomial:
\begin{equation}\label{eq:1.1}  
p(x)=p_0 + p_{1}x+p_2x^2 \cdots +p_{L-1}x^{L-1}+x^L\in \mathbb{F}_2[x]
\end{equation}
is the characteristic polynomial of the recurring sequence and the sequence is generated by $p(x)$. 

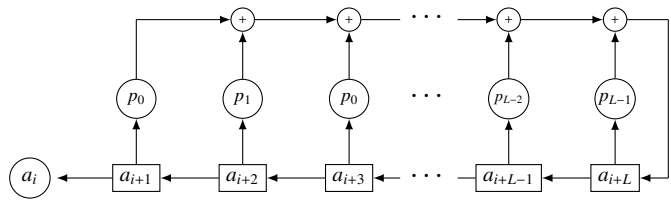
\begin{figure}[t]
\caption{LFSR of length $L$\label{fig:LFSR}}
\begin{center}
\begin{tikzpicture}[scale=0.7]
\draw (0,-1.5) node [draw,scale=0.8](R1){\phantom{$a_{i+1}$}};
\draw (0,-1.5) node [scale=0.8]{$a_{i+1}$};
\draw (2,-1.5) node [draw,scale=0.8](R2){\phantom{$u_{i+2}$}};
\draw (2,-1.5) node [scale=0.8]{$a_{i+2}$};
\draw (4,-1.5) node [ draw,scale=0.8](R3){\phantom{$a_{i+3}$}};
\draw (4,-1.5) node [scale=0.8]{$a_{i+3}$};
\draw (5.5,-1.5) node [scale=1.1](puntos2){$\cdots$};
\draw (7,-1.5) node [ draw,scale=0.8](R4){$a_{i+L-1}$};
\draw (9,-1.5) node [ draw,scale=0.8](R5){$a_{i+L}$};

\draw (0,0) node [draw, circle,scale=0.7](C1){$p_0$};
\draw (2,0) node [draw, circle,scale=0.7](C2){$p_1$};
\draw (4,0) node [draw, circle,scale=0.7](C3){$p_0$};
\draw (5.5,0) node [scale=1.1](puntos){$\cdots$};
\draw (7,0) node [draw, circle,scale=0.7](C4){\phantom{$d_2$}};
\draw (7,0) node [scale=0.6]{$p_{L-2}$};
\draw (9,0) node [,draw, circle,scale=0.7](C5){\phantom{$d_2$}};
\draw (9,0) node [scale=0.7]{$p_{L-1}$};


\draw (2,1.5) node [draw,circle, scale=0.5](S1){$+$};
\draw (4,1.5) node [draw,circle, scale=0.5](S2){$+$};
\draw (5.5,1.5) node [scale=1.1](puntos1){$\cdots$};
\draw (7,1.5) node [draw,circle, scale=0.5](S3){$+$};
\draw (9,1.5) node [draw,circle, scale=0.5](S4){$+$};

\draw (C1)--(0,1.5);
\draw [-latex] (C2)--(S1) ;
\draw [-latex] (C3)--(S2) ;
\draw [-latex] (C4)--(S3) ;
\draw [-latex] (C5)--(S4) ;

\draw [-latex] (0,1.5) --(S1);
\draw [-latex] (S1)--(S2);
\draw  (S2) --(puntos1);
\draw [-latex] (puntos1) --(S3);
\draw [-latex] (S3)--(S4);
\draw (S4)--(10,1.5);
\draw (10,1.5)--(10,-1.5);
\draw [-latex](10,-1.5)--(R5);

\draw [-latex] (R5) --(R4);
\draw (R4)--(puntos2);
\draw [-latex] (puntos2) --(R3);
\draw [-latex] (R3)--(R2);
\draw [-latex](R2)--(R1);
\draw [-latex](R1)--(-1.5,-1.5);

\draw [-latex] (R1)--(C1);
\draw [-latex] (R2)--(C2);
\draw [-latex] (R3)--(C3);
\draw [-latex] (R4)--(C4);
\draw [-latex] (R5)--(C5);



\draw (-2,-1.5)node[scale=0.8,draw,circle]{$a_{i}$};

\end{tikzpicture}
\end{center}
\end{figure}

Traditionally, the Linear Feedback Shift Registers (LFSRs) \cite{Golomb1982bk} implement the linear recurring sequences. In fact, LFSRs are electronic devices whose information units are the elements of $\mathbb{F}_2$. They are made up of $L$ interconnected memory cells (stages) that shift their contents to their next stages and feedback to the empty stage. The register that generates the linear recurring sequence in equation~(\ref{eq:1}) is shown in Figure~\ref{fig:LFSR}. If $p(x)$ is a primitive polynomial \cite{Golomb1982bk}, then the LFSR is called a maximal-period LFSR and generates a PN-sequence (Pseudo Noise sequence) with maximum period of value $T = 2^{L}-1$.

A widely used measure to assess the security of a sequence, especially in potential cryptographic applications, is the linear complexity ($LC$). 
This parameter denotes the length of the shortest Linear Feedback Shift Register (LFSR) capable of generating the sequence. Essentially, the $LC$ of a sequence corresponds to the lowest order of its linear recurrence relationship.

In cryptographic terms, the $LC$ determines the segment of the sequence that needs to be intercepted to recover the remaining bits. Larger values of $LC$ are preferable for heightened security. 
In modern cryptographic applications, it is essential to generate sequences with extremely long periods to ensure high security levels and resistance to attacks such as brute-force or time-memory trade-offs. A commonly accepted benchmark is a period of at least $2^{128}$ bits, which matches the key length used in widely adopted encryption standards like AES-$128$ \cite{Paar2024bk}.

Now we are ready to introduce one of the main concepts of this work. 

\begin{definition}

We say that the sequence $\{s_j\}$
 is obtained interleaving the sequences
 $\{u^{(1)}_i\}$, $\{u^{(2)}_i\}$, $\ldots$, $\{u^{(t)}_i\}$, all of them of period $T$, if it has the following form:
 $$\{s_j\}=\left\{u_0^{(1)},u_0^{(2)},\ldots, u_0^{(t)}, u_1^{(1)},u_1^{(2)},\ldots, u_1^{(t)}, \ldots, u_{T-1}^{(1)},u_{T-1}^{(2)},\ldots, u_{T-1}^{(t)} \right\}.$$
 We call this sequence a \textbf{$t$-interleaving sequence}.
\end{definition}

In \cite{Cardell2021a}, the authors showed that when we interleave $t$ shifted versions of the same PN-sequence, the resultant $t$-interleaving sequence has linear complexity $LC\leq t\cdot L$.
Given a fixed primitive polynomial, when interleaving shifted versions of one PN-sequence, they showed that in almost 90\% of the cases, we have that $LC$ reaches the maximum value $t\cdot L$.
In some cases, depending on the shifts,  the complexity is of the form $s\cdot L$, with $s=1,2, \ldots, t-1$.
They conducted an initial analysis on the randomness of these sequences, revealing a notable finding: nearly all sequences exhibit robust cryptographic properties (see \cite{Cardell2021a} for more details).

In this work we only consider  the cases with maximum $LC$.

\subsection{Generalities of cellular automata}\label{sec:binomial}
 
\textbf{Cellular automata} (CA) are devices composed of a finite number of cells whose content is updated according to a \emph{rule} or function with $k$ variables \cite{Das1990}.
The   cell in position $i$ at time $t+1$, denoted by  $x_{i}^{t+1}$, depends on the state of the $k$ neighbour cells at time $t$.
If these rules are composed exclusively of XOR operations, then the CA are \textbf{linear}.
We can distinguish our CA between  \textbf{regular} (every cell  follows the same rule) or \textbf{hybrid} (different rules) and    \textbf{cyclic} (extreme cells are adjacent) or \textbf{null} (extreme cells are considered adjacent to zero columns).
In this work, we prove that the interleaving sequences can be generated by two families of linear cellular automata: one consisting of regular/cyclic CAs, and the other composed of null/hybrid CAs.

For $k=3$, rules 102, 60, 150 and 90 are given by:
 \medskip

\begin{minipage}[]{0.4\textwidth}
\centering
\resizebox{6cm}{!} {
$
\begin{array}{c}
\text{\textbf{Rule 102:} }x_{i}^{t+1}=x_{i}^{t}+x_{i+1}^{t}\\
\\
\begin{array}{|c|c|c|c|c|c|c|c|}\hline
111 & 110 & 101 & 100 & 011 & 010 & 001 & 000\\ \hline
0  &  1  &  1  & 0   & 0   &  1  &  1  & 0 \\ \hline
\end{array}\\
\end{array}
$
}
\end{minipage}
\begin{minipage}[]{0.5\textwidth}
\centering
\resizebox{6cm}{!} {
$
\begin{array}{c}
\text{\textbf{Rule 60:} }x_{i}^{t+1}=x_{i-1}^{t}+x_{i}^{t}\\
\\
\begin{array}{|c|c|c|c|c|c|c|c|}\hline
111 & 110 & 101 & 100 & 011 & 010 & 001 & 000\\ \hline
0  &  0  &  1  & 1   & 1   &  1  &  0  & 0 \\ \hline
\end{array}\\
\end{array}
$
}
\end{minipage}
\bigskip

\begin{minipage}[]{0.4\textwidth}
\centering
\resizebox{6cm}{!} {
$
\begin{array}{c}
\text{\textbf{Rule 150:} }x_{i}^{t+1}=x_{i-1}^{t}+x_{i}^{t}+x_{i+1}^{t}\\
\\
\begin{array}{|c|c|c|c|c|c|c|c|}\hline
111 & 110 & 101 & 100 & 011 & 010 & 001 & 000\\ \hline
1  &  0  &  0  & 1   & 0   &  1  &  1  & 0 \\ \hline
\end{array}\\
\end{array}
$
}
\end{minipage}
\begin{minipage}[]{0.5\textwidth}
\centering
\resizebox{6cm}{!} {
$
\begin{array}{c}
\text{\textbf{Rule 90:} }x_{i}^{t+1}=x_{i-1}^{t}+x_{i+1}^{t}\\
\\
\begin{array}{|c|c|c|c|c|c|c|c|}\hline
111 & 110 & 101 & 100 & 011 & 010 & 001 & 000\\ \hline
0  &  1  &  0  & 1   & 1   &  0  &  1  & 0 \\ \hline
\end{array}\\
\end{array}
$
}
\end{minipage}

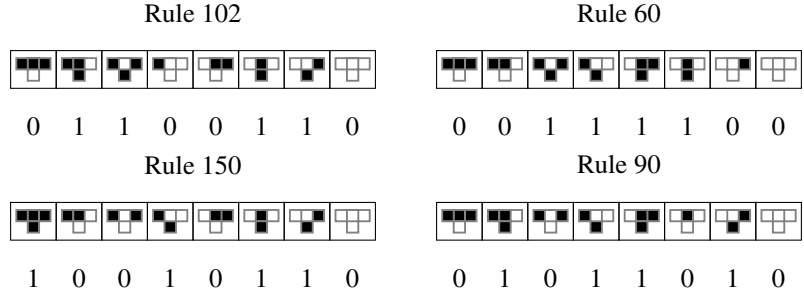
\begin{figure}
\caption{Rules 102, 60, 150 and 90 seen with Wolfram's notation \label{fig:1}}
\centering

\begin{tikzpicture}[scale=0.5]
\draw (4.8,2) node {Rule 102};
\draw (0.6,-1) node{0};
\draw (1.8,-1) node{1};
\draw (3,-1) node{1};
\draw (4.2,-1) node{0};
\draw (5.4,-1) node{0};
\draw (6.6,-1) node{1};
\draw (7.8,-1) node{1};
\draw (9,-1) node{0};

\draw (0,0) rectangle (1.2,1);
\draw [color=white, fill=black](0.15,0.5)rectangle(0.45,0.8);
\draw [color=white, fill=black](0.45,0.5)rectangle(0.75,0.8);
\draw [color=white, fill=black](0.75,0.5)rectangle(1.05,0.8);
\draw [color=white](0.45,0.2)rectangle(0.75,0.5);
\draw [color=gray!100, line width=0.7](0.45,0.8)--(0.15,0.8)--(0.15,0.5)--(1.05,0.5)--(1.05,0.8)--(0.45,0.8)--(0.45,0.2)-- (0.75,0.2)-- (0.75,0.8)-- (0.45,0.8);
\draw (1.2,0) rectangle (2.4,1);
\draw [color=white, fill=black](1.35,0.5)rectangle(1.65,0.8);
\draw [color=white, fill=black](1.65,0.5)rectangle(1.95,0.8);
\draw [color=white](1.95,0.5)rectangle(2.25,0.8);
\draw [color=white,fill=black](1.65,0.2)rectangle(1.95,0.5);
\draw [color=gray!100, line width=0.7](1.65,0.8)--(1.35,0.8)--(1.35,0.5)--(2.25,0.5)--(2.25,0.8)--(1.65,0.8)--(1.65,0.2)-- (1.95,0.2)-- (1.95,0.8)-- (1.65,0.8);
\draw (2.4,0) rectangle (3.6,1);
\draw [color=white, fill=black](2.55,0.5)rectangle(2.85,0.8);
\draw [color=white](2.85,0.5)rectangle(3.15,0.8);
\draw [color=white, fill=black](3.15,0.5)rectangle(3.45,0.8);
\draw [color=white, fill=black](2.85,0.2)rectangle(3.15,0.5);
\draw [color=gray!100, line width=0.7](2.85,0.8)--(2.55,0.8)--(2.55,0.5)--(3.45,0.5)--(3.45,0.8)--(2.85,0.8)--(2.85,0.2)-- (3.15,0.2)-- (3.15,0.8)-- (2.85,0.8);
\draw (3.6,0) rectangle (4.8,1);
\draw [color=white, fill=black](3.75,0.5)rectangle(4.05,0.8);
\draw [color=white](4.05,0.5)rectangle(4.35,0.8);
\draw [color=white](4.35,0.5)rectangle(4.65,0.8);
\draw [color=white](4.05,0.2)rectangle(4.35,0.5);
\draw [color=gray!100, line width=0.7](4.05,0.8)--(3.75,0.8)--(3.75,0.5)--(4.65,0.5)--(4.65,0.8)--(4.05,0.8)--(4.05,0.2)-- (4.35,0.2)-- (4.35,0.8)-- (4.05,0.8);
\draw (4.8,0) rectangle (6,1);
\draw [color=white](4.95,0.5)rectangle(5.25,0.8);
\draw [color=white, fill=black](5.25,0.5)rectangle(5.55,0.8);
\draw [color=white, fill=black](5.55,0.5)rectangle(5.85,0.8);
\draw [color=white](5.25,0.2)rectangle(5.55,0.5);
\draw [color=gray!100, line width=0.7](5.25,0.8)--(4.95,0.8)--(4.95,0.5)--(5.85,0.5)--(5.85,0.8)--(5.25,0.8)--(5.25,0.2)-- (5.55,0.2)-- (5.55,0.8)-- (5.25,0.8);
\draw (6,0) rectangle (7.2,1);
\draw [color=white](6.15,0.5)rectangle(6.45,0.8);
\draw [color=white, fill=black](6.45,0.5)rectangle(6.75,0.8);
\draw [color=white](6.75,0.5)rectangle(7.05,0.8);
\draw [color=white, fill=black](6.45,0.2)rectangle(6.75,0.5);
\draw [color=gray!100, line width=0.7](6.45,0.8)--(6.15,0.8)--(6.15,0.5)--(7.05,0.5)--(7.05,0.8)--(6.45,0.8)--(6.45,0.2)-- (6.75,0.2)-- (6.75,0.8)-- (6.45,0.8);
\draw (7.2,0) rectangle (8.4,1);
\draw [color=white](7.35,0.5)rectangle(7.65,0.8);
\draw [color=white](7.65,0.5)rectangle(7.95,0.8);
\draw [color=white, fill=black](7.95,0.5)rectangle(8.25,0.8);
\draw [color=white,fill=black](7.65,0.2)rectangle(7.95,0.5);
\draw [color=gray!100, line width=0.7](7.65,0.8)--(7.35,0.8)--(7.35,0.5)--(8.25,0.5)--(8.25,0.8)--(7.65,0.8)--(7.65,0.2)-- (7.95,0.2)-- (7.95,0.8)-- (7.65,0.8);
\draw (8.4,0) rectangle (9.6,1);
\draw [color=white](8.55,0.5)rectangle(8.85,0.8);
\draw [color=white](8.85,0.5)rectangle(9.15,0.8);
\draw [color=white](9.15,0.5)rectangle(9.45,0.8);
\draw [color=white](8.85,0.2)rectangle(9.15,0.5);
\draw [color=gray!100, line width=0.7](8.85,0.8)--(8.55,0.8)--(8.55,0.5)--(9.45,0.5)--(9.45,0.8)--(8.85,0.8)--(8.85,0.2)-- (9.15,0.2)-- (9.15,0.8)-- (8.85,0.8);
\end{tikzpicture}
\qquad
\begin{tikzpicture}[scale=0.5]
\draw (4.8,2) node {Rule 60};
\draw (0.6,-1) node{0};
\draw (1.8,-1) node{0};
\draw (3,-1) node{1};
\draw (4.2,-1) node{1};
\draw (5.4,-1) node{1};
\draw (6.6,-1) node{1};
\draw (7.8,-1) node{0};
\draw (9,-1) node{0};

\draw (0,0) rectangle (1.2,1);
\draw [color=white, fill=black](0.15,0.5)rectangle(0.45,0.8);
\draw [color=white, fill=black](0.45,0.5)rectangle(0.75,0.8);
\draw [color=white, fill=black](0.75,0.5)rectangle(1.05,0.8);
\draw [color=white](0.45,0.2)rectangle(0.75,0.5);
\draw [color=gray!100, line width=0.7](0.45,0.8)--(0.15,0.8)--(0.15,0.5)--(1.05,0.5)--(1.05,0.8)--(0.45,0.8)--(0.45,0.2)-- (0.75,0.2)-- (0.75,0.8)-- (0.45,0.8);
\draw (1.2,0) rectangle (2.4,1);
\draw [color=white, fill=black](1.35,0.5)rectangle(1.65,0.8);
\draw [color=white, fill=black](1.65,0.5)rectangle(1.95,0.8);
\draw [color=white](1.95,0.5)rectangle(2.25,0.8);
\draw [color=white](1.65,0.2)rectangle(1.95,0.5);
\draw [color=gray!100, line width=0.7](1.65,0.8)--(1.35,0.8)--(1.35,0.5)--(2.25,0.5)--(2.25,0.8)--(1.65,0.8)--(1.65,0.2)-- (1.95,0.2)-- (1.95,0.8)-- (1.65,0.8);
\draw (2.4,0) rectangle (3.6,1);
\draw [color=white, fill=black](2.55,0.5)rectangle(2.85,0.8);
\draw [color=white](2.85,0.5)rectangle(3.15,0.8);
\draw [color=white, fill=black](3.15,0.5)rectangle(3.45,0.8);
\draw [color=white, fill=black](2.85,0.2)rectangle(3.15,0.5);
\draw [color=gray!100, line width=0.7](2.85,0.8)--(2.55,0.8)--(2.55,0.5)--(3.45,0.5)--(3.45,0.8)--(2.85,0.8)--(2.85,0.2)-- (3.15,0.2)-- (3.15,0.8)-- (2.85,0.8);
\draw (3.6,0) rectangle (4.8,1);
\draw [color=white, fill=black](3.75,0.5)rectangle(4.05,0.8);
\draw [color=white](4.05,0.5)rectangle(4.35,0.8);
\draw [color=white](4.35,0.5)rectangle(4.65,0.8);
\draw [color=white, fill=black](4.05,0.2)rectangle(4.35,0.5);
\draw [color=gray!100, line width=0.7](4.05,0.8)--(3.75,0.8)--(3.75,0.5)--(4.65,0.5)--(4.65,0.8)--(4.05,0.8)--(4.05,0.2)-- (4.35,0.2)-- (4.35,0.8)-- (4.05,0.8);
\draw (4.8,0) rectangle (6,1);
\draw [color=white](4.95,0.5)rectangle(5.25,0.8);
\draw [color=white, fill=black](5.25,0.5)rectangle(5.55,0.8);
\draw [color=white, fill=black](5.55,0.5)rectangle(5.85,0.8);
\draw [color=white, fill=black](5.25,0.2)rectangle(5.55,0.5);
\draw [color=gray!100, line width=0.7](5.25,0.8)--(4.95,0.8)--(4.95,0.5)--(5.85,0.5)--(5.85,0.8)--(5.25,0.8)--(5.25,0.2)-- (5.55,0.2)-- (5.55,0.8)-- (5.25,0.8);
\draw (6,0) rectangle (7.2,1);
\draw [color=white](6.15,0.5)rectangle(6.45,0.8);
\draw [color=white, fill=black](6.45,0.5)rectangle(6.75,0.8);
\draw [color=white](6.75,0.5)rectangle(7.05,0.8);
\draw [color=white, fill=black](6.45,0.2)rectangle(6.75,0.5);
\draw [color=gray!100, line width=0.7](6.45,0.8)--(6.15,0.8)--(6.15,0.5)--(7.05,0.5)--(7.05,0.8)--(6.45,0.8)--(6.45,0.2)-- (6.75,0.2)-- (6.75,0.8)-- (6.45,0.8);
\draw (7.2,0) rectangle (8.4,1);
\draw [color=white](7.35,0.5)rectangle(7.65,0.8);
\draw [color=white](7.65,0.5)rectangle(7.95,0.8);
\draw [color=white, fill=black](7.95,0.5)rectangle(8.25,0.8);
\draw [color=white](7.65,0.2)rectangle(7.95,0.5);
\draw [color=gray!100, line width=0.7](7.65,0.8)--(7.35,0.8)--(7.35,0.5)--(8.25,0.5)--(8.25,0.8)--(7.65,0.8)--(7.65,0.2)-- (7.95,0.2)-- (7.95,0.8)-- (7.65,0.8);
\draw (8.4,0) rectangle (9.6,1);
\draw [color=white](8.55,0.5)rectangle(8.85,0.8);
\draw [color=white](8.85,0.5)rectangle(9.15,0.8);
\draw [color=white](9.15,0.5)rectangle(9.45,0.8);
\draw [color=white](8.85,0.2)rectangle(9.15,0.5);
\draw [color=gray!100, line width=0.7](8.85,0.8)--(8.55,0.8)--(8.55,0.5)--(9.45,0.5)--(9.45,0.8)--(8.85,0.8)--(8.85,0.2)-- (9.15,0.2)-- (9.15,0.8)-- (8.85,0.8);
\end{tikzpicture}

\begin{tikzpicture}[scale=0.5]
\draw (4.8,2) node {Rule 150};
\draw (0.6,-1) node{1};
\draw (1.8,-1) node{0};
\draw (3,-1) node{0};
\draw (4.2,-1) node{1};
\draw (5.4,-1) node{0};
\draw (6.6,-1) node{1};
\draw (7.8,-1) node{1};
\draw (9,-1) node{0};

\draw (0,0) rectangle (1.2,1);
\draw [color=white, fill=black](0.15,0.5)rectangle(0.45,0.8);
\draw [color=white, fill=black](0.45,0.5)rectangle(0.75,0.8);
\draw [color=white, fill=black](0.75,0.5)rectangle(1.05,0.8);
\draw [color=white, fill=black](0.45,0.2)rectangle(0.75,0.5);
\draw [color=gray!100, line width=0.7](0.45,0.8)--(0.15,0.8)--(0.15,0.5)--(1.05,0.5)--(1.05,0.8)--(0.45,0.8)--(0.45,0.2)-- (0.75,0.2)-- (0.75,0.8)-- (0.45,0.8);
\draw (1.2,0) rectangle (2.4,1);
\draw [color=white, fill=black](1.35,0.5)rectangle(1.65,0.8);
\draw [color=white, fill=black](1.65,0.5)rectangle(1.95,0.8);
\draw [color=white](1.95,0.5)rectangle(2.25,0.8);
\draw [color=white](1.65,0.2)rectangle(1.95,0.5);
\draw [color=gray!100, line width=0.7](1.65,0.8)--(1.35,0.8)--(1.35,0.5)--(2.25,0.5)--(2.25,0.8)--(1.65,0.8)--(1.65,0.2)-- (1.95,0.2)-- (1.95,0.8)-- (1.65,0.8);
\draw (2.4,0) rectangle (3.6,1);
\draw [color=white, fill=black](2.55,0.5)rectangle(2.85,0.8);
\draw [color=white](2.85,0.5)rectangle(3.15,0.8);
\draw [color=white, fill=black](3.15,0.5)rectangle(3.45,0.8);
\draw [color=white](2.85,0.2)rectangle(3.15,0.5);
\draw [color=gray!100, line width=0.7](2.85,0.8)--(2.55,0.8)--(2.55,0.5)--(3.45,0.5)--(3.45,0.8)--(2.85,0.8)--(2.85,0.2)-- (3.15,0.2)-- (3.15,0.8)-- (2.85,0.8);
\draw (3.6,0) rectangle (4.8,1);
\draw [color=white, fill=black](3.75,0.5)rectangle(4.05,0.8);
\draw [color=white](4.05,0.5)rectangle(4.35,0.8);
\draw [color=white](4.35,0.5)rectangle(4.65,0.8);
\draw [color=white, fill=black](4.05,0.2)rectangle(4.35,0.5);
\draw [color=gray!100, line width=0.7](4.05,0.8)--(3.75,0.8)--(3.75,0.5)--(4.65,0.5)--(4.65,0.8)--(4.05,0.8)--(4.05,0.2)-- (4.35,0.2)-- (4.35,0.8)-- (4.05,0.8);
\draw (4.8,0) rectangle (6,1);
\draw [color=white](4.95,0.5)rectangle(5.25,0.8);
\draw [color=white, fill=black](5.25,0.5)rectangle(5.55,0.8);
\draw [color=white, fill=black](5.55,0.5)rectangle(5.85,0.8);
\draw [color=white](5.25,0.2)rectangle(5.55,0.5);
\draw [color=gray!100, line width=0.7](5.25,0.8)--(4.95,0.8)--(4.95,0.5)--(5.85,0.5)--(5.85,0.8)--(5.25,0.8)--(5.25,0.2)-- (5.55,0.2)-- (5.55,0.8)-- (5.25,0.8);
\draw (6,0) rectangle (7.2,1);
\draw [color=white](6.15,0.5)rectangle(6.45,0.8);
\draw [color=white, fill=black](6.45,0.5)rectangle(6.75,0.8);
\draw [color=white](6.75,0.5)rectangle(7.05,0.8);
\draw [color=white, fill=black](6.45,0.2)rectangle(6.75,0.5);
\draw [color=gray!100, line width=0.7](6.45,0.8)--(6.15,0.8)--(6.15,0.5)--(7.05,0.5)--(7.05,0.8)--(6.45,0.8)--(6.45,0.2)-- (6.75,0.2)-- (6.75,0.8)-- (6.45,0.8);
\draw (7.2,0) rectangle (8.4,1);
\draw [color=white](7.35,0.5)rectangle(7.65,0.8);
\draw [color=white](7.65,0.5)rectangle(7.95,0.8);
\draw [color=white, fill=black](7.95,0.5)rectangle(8.25,0.8);
\draw [color=white, fill=black](7.65,0.2)rectangle(7.95,0.5);
\draw [color=gray!100, line width=0.7](7.65,0.8)--(7.35,0.8)--(7.35,0.5)--(8.25,0.5)--(8.25,0.8)--(7.65,0.8)--(7.65,0.2)-- (7.95,0.2)-- (7.95,0.8)-- (7.65,0.8);
\draw (8.4,0) rectangle (9.6,1);
\draw [color=white](8.55,0.5)rectangle(8.85,0.8);
\draw [color=white](8.85,0.5)rectangle(9.15,0.8);
\draw [color=white](9.15,0.5)rectangle(9.45,0.8);
\draw [color=white](8.85,0.2)rectangle(9.15,0.5);
\draw [color=gray!100, line width=0.7](8.85,0.8)--(8.55,0.8)--(8.55,0.5)--(9.45,0.5)--(9.45,0.8)--(8.85,0.8)--(8.85,0.2)-- (9.15,0.2)-- (9.15,0.8)-- (8.85,0.8);
\end{tikzpicture}
\qquad
\begin{tikzpicture}[scale=0.5]
\draw (4.8,2) node {Rule 90};
\draw (0.6,-1) node{0};
\draw (1.8,-1) node{1};
\draw (3,-1) node{0};
\draw (4.2,-1) node{1};
\draw (5.4,-1) node{1};
\draw (6.6,-1) node{0};
\draw (7.8,-1) node{1};
\draw (9,-1) node{0};

\draw (0,0) rectangle (1.2,1);
\draw [color=white, fill=black](0.15,0.5)rectangle(0.45,0.8);
\draw [color=white, fill=black](0.45,0.5)rectangle(0.75,0.8);
\draw [color=white, fill=black](0.75,0.5)rectangle(1.05,0.8);
\draw [color=white](0.45,0.2)rectangle(0.75,0.5);
\draw [color=gray!100, line width=0.7](0.45,0.8)--(0.15,0.8)--(0.15,0.5)--(1.05,0.5)--(1.05,0.8)--(0.45,0.8)--(0.45,0.2)-- (0.75,0.2)-- (0.75,0.8)-- (0.45,0.8);
\draw (1.2,0) rectangle (2.4,1);
\draw [color=white, fill=black](1.35,0.5)rectangle(1.65,0.8);
\draw [color=white, fill=black](1.65,0.5)rectangle(1.95,0.8);
\draw [color=white](1.95,0.5)rectangle(2.25,0.8);
\draw [color=white, fill=black](1.65,0.2)rectangle(1.95,0.5);
\draw [color=gray!100, line width=0.7](1.65,0.8)--(1.35,0.8)--(1.35,0.5)--(2.25,0.5)--(2.25,0.8)--(1.65,0.8)--(1.65,0.2)-- (1.95,0.2)-- (1.95,0.8)-- (1.65,0.8);
\draw (2.4,0) rectangle (3.6,1);
\draw [color=white, fill=black](2.55,0.5)rectangle(2.85,0.8);
\draw [color=white](2.85,0.5)rectangle(3.15,0.8);
\draw [color=white, fill=black](3.15,0.5)rectangle(3.45,0.8);
\draw [color=white](2.85,0.2)rectangle(3.15,0.5);
\draw [color=gray!100, line width=0.7](2.85,0.8)--(2.55,0.8)--(2.55,0.5)--(3.45,0.5)--(3.45,0.8)--(2.85,0.8)--(2.85,0.2)-- (3.15,0.2)-- (3.15,0.8)-- (2.85,0.8);
\draw (3.6,0) rectangle (4.8,1);
\draw [color=white, fill=black](3.75,0.5)rectangle(4.05,0.8);
\draw [color=white](4.05,0.5)rectangle(4.35,0.8);
\draw [color=white](4.35,0.5)rectangle(4.65,0.8);
\draw [color=white, fill=black](4.05,0.2)rectangle(4.35,0.5);
\draw [color=gray!100, line width=0.7](4.05,0.8)--(3.75,0.8)--(3.75,0.5)--(4.65,0.5)--(4.65,0.8)--(4.05,0.8)--(4.05,0.2)-- (4.35,0.2)-- (4.35,0.8)-- (4.05,0.8);
\draw (4.8,0) rectangle (6,1);
\draw [color=white](4.95,0.5)rectangle(5.25,0.8);
\draw [color=white, fill=black](5.25,0.5)rectangle(5.55,0.8);
\draw [color=white, fill=black](5.55,0.5)rectangle(5.85,0.8);
\draw [color=white, fill=black](5.25,0.2)rectangle(5.55,0.5);
\draw [color=gray!100, line width=0.7](5.25,0.8)--(4.95,0.8)--(4.95,0.5)--(5.85,0.5)--(5.85,0.8)--(5.25,0.8)--(5.25,0.2)-- (5.55,0.2)-- (5.55,0.8)-- (5.25,0.8);
\draw (6,0) rectangle (7.2,1);
\draw [color=white](6.15,0.5)rectangle(6.45,0.8);
\draw [color=white, fill=black](6.45,0.5)rectangle(6.75,0.8);
\draw [color=white](6.75,0.5)rectangle(7.05,0.8);
\draw [color=white](6.45,0.2)rectangle(6.75,0.5);
\draw [color=gray!100, line width=0.7](6.45,0.8)--(6.15,0.8)--(6.15,0.5)--(7.05,0.5)--(7.05,0.8)--(6.45,0.8)--(6.45,0.2)-- (6.75,0.2)-- (6.75,0.8)-- (6.45,0.8);
\draw (7.2,0) rectangle (8.4,1);
\draw [color=white](7.35,0.5)rectangle(7.65,0.8);
\draw [color=white](7.65,0.5)rectangle(7.95,0.8);
\draw [color=white, fill=black](7.95,0.5)rectangle(8.25,0.8);
\draw [color=white, fill=black](7.65,0.2)rectangle(7.95,0.5);
\draw [color=gray!100, line width=0.7](7.65,0.8)--(7.35,0.8)--(7.35,0.5)--(8.25,0.5)--(8.25,0.8)--(7.65,0.8)--(7.65,0.2)-- (7.95,0.2)-- (7.95,0.8)-- (7.65,0.8);
\draw (8.4,0) rectangle (9.6,1);
\draw [color=white](8.55,0.5)rectangle(8.85,0.8);
\draw [color=white](8.85,0.5)rectangle(9.15,0.8);
\draw [color=white](9.15,0.5)rectangle(9.45,0.8);
\draw [color=white](8.85,0.2)rectangle(9.15,0.5);
\draw [color=gray!100, line width=0.7](8.85,0.8)--(8.55,0.8)--(8.55,0.5)--(9.45,0.5)--(9.45,0.8)--(8.85,0.8)--(8.85,0.2)-- (9.15,0.2)-- (9.15,0.8)-- (8.85,0.8);
\end{tikzpicture}
\end{figure}

\begin{figure}[t]
\caption{AC-images generated with rules 102, 60, 150 and 90 (top-down time space diagrams)\label{fig:2}}
\centering
\begin{tikzpicture}[scale=1.25]
\draw (-1.6,1.9 )node{Rule 102};
\begin{scope}[x=-1cm]
\draw [fill=black] (1.6,0)--(3.2,0)--(3.2,0.1)--(3.1,0.1)--(3.1,0.2)--(3,0.2)--(3,0.3)--(2.9,0.3)--(2.9,0.4)--(2.8,0.4)--(2.8,0.5)--(2.7,0.5)--(2.7,0.6)--(2.6,0.6)--(2.6,0.7)--(2.5,0.7)--(2.5,0.8)--(2.4,0.8)--(2.4,0.9)--(2.3,0.9)--(2.3,1)--(2.2,1)--(2.2,1.1)--(2.1,1.1)--(2.1,1.2)--(2,1.2)--(2,1.3)--(1.9,1.3)--(1.9,1.4)--(1.8,1.4)--(1.8,1.5)--(1.7,1.5)--(1.7,1.6)--(1.6,1.6)--cycle;
\draw  [fill=white](1.7,0.8)--(2.4,0.8)--(2.4,0.1)--(2.3,0.1)--(2.3,0.2)--(2.2,0.2)--(2.2,0.3)--(2.1,0.3)--(2.1,0.4)--(2.0,0.4)--(2.0,0.5)--(1.9,0.5)--(1.9,0.6)--(1.8,0.6)--(1.8,0.7)--(1.7,0.7)--cycle;
\draw  [fill=white](1.7,1.2)--(2,1.2)--(2,0.9)--(1.9,0.9)--(1.9,1)--(1.8,1)--(1.8,1.1)--(1.7,1.1)--cycle;
\draw  [fill=white](2.5,0.4)--(2.8,0.4)--(2.8,0.1)--(2.7,0.1)--(2.7,0.2)--(2.6,0.2)--(2.6,0.3)--(2.5,0.3)--cycle;

\draw  [fill=white](1.7,0.4)--(2,0.4)--(2,0.1)--(1.9,0.1)--(1.9,0.2)--(1.8,0.2)--(1.8,0.3)--(1.7,0.3)--cycle;

\draw[fill=white](1.7,0.1)rectangle(1.8,0.2);
\draw[fill=white](1.7,0.5)rectangle(1.8,0.6);
\draw[fill=white](1.7,0.9)rectangle(1.8,1.0);
\draw[fill=white](1.7,1.3)rectangle(1.8,1.4);

\draw[fill=white](2.1,0.1)rectangle(2.2,0.2);
\draw[fill=white](2.1,0.9)rectangle(2.2,1);

\draw[fill=white](2.5,0.1)rectangle(2.6,0.2);
\draw[fill=white](2.5,0.5)rectangle(2.6,0.6);

\draw[fill=white](2.9,0.1)rectangle(3,0.2);
\end{scope}
\draw [step=0.1, color=gray!100] (-3.2,0)grid(0,1.6);
\draw [  color=gray!100](-3.2,0)--(-3.2,1.6);
\end{tikzpicture}
\qquad
\begin{tikzpicture}[scale=1.25]
\draw (1.6,1.9 )node{Rule 60};
\draw [fill=black] (1.6,0)--(3.2,0)--(3.2,0.1)--(3.1,0.1)--(3.1,0.2)--(3,0.2)--(3,0.3)--(2.9,0.3)--(2.9,0.4)--(2.8,0.4)--(2.8,0.5)--(2.7,0.5)--(2.7,0.6)--(2.6,0.6)--(2.6,0.7)--(2.5,0.7)--(2.5,0.8)--(2.4,0.8)--(2.4,0.9)--(2.3,0.9)--(2.3,1)--(2.2,1)--(2.2,1.1)--(2.1,1.1)--(2.1,1.2)--(2,1.2)--(2,1.3)--(1.9,1.3)--(1.9,1.4)--(1.8,1.4)--(1.8,1.5)--(1.7,1.5)--(1.7,1.6)--(1.6,1.6)--cycle;
\draw  [fill=white](1.7,0.8)--(2.4,0.8)--(2.4,0.1)--(2.3,0.1)--(2.3,0.2)--(2.2,0.2)--(2.2,0.3)--(2.1,0.3)--(2.1,0.4)--(2.0,0.4)--(2.0,0.5)--(1.9,0.5)--(1.9,0.6)--(1.8,0.6)--(1.8,0.7)--(1.7,0.7)--cycle;
\draw  [fill=white](1.7,1.2)--(2,1.2)--(2,0.9)--(1.9,0.9)--(1.9,1)--(1.8,1)--(1.8,1.1)--(1.7,1.1)--cycle;
\draw  [fill=white](2.5,0.4)--(2.8,0.4)--(2.8,0.1)--(2.7,0.1)--(2.7,0.2)--(2.6,0.2)--(2.6,0.3)--(2.5,0.3)--cycle;
\draw  [fill=white](1.7,0.4)--(2,0.4)--(2,0.1)--(1.9,0.1)--(1.9,0.2)--(1.8,0.2)--(1.8,0.3)--(1.7,0.3)--cycle;

\draw[fill=white](1.7,0.1)rectangle(1.8,0.2);
\draw[fill=white](1.7,0.5)rectangle(1.8,0.6);
\draw[fill=white](1.7,0.9)rectangle(1.8,1.0);
\draw[fill=white](1.7,1.3)rectangle(1.8,1.4);

\draw[fill=white](2.1,0.1)rectangle(2.2,0.2);
\draw[fill=white](2.1,0.9)rectangle(2.2,1);

\draw[fill=white](2.5,0.1)rectangle(2.6,0.2);
\draw[fill=white](2.5,0.5)rectangle(2.6,0.6);

\draw[fill=white](2.9,0.1)rectangle(3,0.2);
\draw [step=0.1, color=gray!100] (0.1,0)grid(3.2,1.6);
\end{tikzpicture}

\vspace{1cm}

\begin{tikzpicture}[scale=1.25]
\draw (1.6,1.9 )node{Rule 150};
\draw [fill=black] (0,0)--(0.2,0)--(0.2,0.2)--(0.1,0.2)--(0.1,0.1)--(0,0.1)--cycle;
\draw [fill=black] (0.4,0.4)--(0.6,0.4)--(0.6,0.6)--(0.5,0.6)--(0.5,0.5)--(0.4,0.5)--cycle;
\draw [fill=black] (0.8,0.8)--(1,0.8)--(1,1)--(0.9,1)--(0.9,0.9)--(0.8,0.9)--cycle;
\draw [fill=black] (1.2,1.2)--(1.4,1.2)--(1.4,1.4)--(1.3,1.4)--(1.3,1.3)--(1.2,1.3)--cycle;

\draw [fill=black](0.9,0)--(1.1,0)--(1.1,0.1)--(1,0.1)--(1,0.2)--(0.9,0.2)--cycle;
\draw [fill=black](0.9,0.4)--(1.1,0.4)--(1.1,0.5)--(1,0.5)--(1,0.6)--(0.9,0.6)--cycle;
\draw [fill=black] (1.2,0)--(1.4,0)--(1.4,0.2)--(1.3,0.2)--(1.3,0.1)--(1.2,0.1)--cycle;
\draw [fill=black] (1.2,0.4)--(1.4,0.4)--(1.4,0.6)--(1.3,0.6)--(1.3,0.5)--(1.2,0.5)--cycle;

\draw [fill=black] (0.3,0)--(0.5,0)--(0.5,0.1)--(0.4,0.1)--(0.4,0.2)--(0.5,0.2)--(0.5,0.3)--(0.4,0.3)--(0.4,0.4)--(0.3,0.4)--(0.3,0.3)--(0.2,0.3)--(0.2,0.2)--(0.3, 0.2)--cycle;

\draw [fill=black] (1.1,0.8)--(1.3,0.8)--(1.3,0.9)--(1.2,0.9)--(1.2,1)--(1.3,1)--(1.3,1.1)--(1.2,1.1)--(1.2,1.2)--(1.1,1.2)--(1.1,1.1)--(1,1.1)--(1,1)--(1.1, 1)--cycle;

\draw [fill=black] (0.6,0)--(0.8,0)--(0.8,0.2)--(0.9,0.2)--(0.9,0.3)--(0.8,0.3)--(0.8,0.6)--(0.9,0.6)--(0.9,0.7)--(0.8,0.7)--(0.8,0.8)--(0.7,0.8)--(0.7,0.7)--(0.6, 0.7)--(0.6,0.6)--(0.7,0.6)--(0.7, 0.3)--(0.6,0.3)--(0.6,0.2)--(0.7,0.2)--(0.7,0.1)--(0.6,0.1)--cycle;

\draw [fill=black](1.5,0)rectangle(1.6,1.6);
\draw [fill=black](1.4,0.2)rectangle(1.5,0.3);
\draw [fill=black](1.4,0.6)rectangle(1.5,0.7);
\draw [fill=black](1.4,0.8)rectangle(1.5,0.9);
\draw [fill=black](1.4,1)rectangle(1.5,1.1);
\draw [fill=black](1.4,1.4)rectangle(1.5,1.5);

\draw [fill=black] (3.1,0)--(2.9,0)--(2.9,0.2)--(3 ,0.2)--(3 ,0.1)--(3.1,0.1)--cycle;
\draw [fill=black] (2.7,0.4)--(2.5,0.4)--(2.5,0.6)--(2.6,0.6)--(2.6,0.5)--(2.7,0.5)--cycle;
\draw [fill=black] (2.3,0.8)--(2.1,0.8)--(2.1,1)--(2.2,1)--(2.2,0.9)--(2.3,0.9)--cycle;
\draw [fill=black] (1.9,1.2)--(1.7,1.2)--(1.7,1.4)--(1.8,1.4)--(1.8,1.3)--(1.9,1.3)--cycle;

\draw [fill=black](2.2,0)--(2,0)--(2,0.1)--(2.1,0.1)--(2.1,0.2)--(2.2,0.2)--cycle;
\draw [fill=black](2.2,0.4)--(2.0,0.4)--(2.0,0.5)--(2.1,0.5)--(2.1,0.6)--(2.2,0.6)--cycle;
\draw [fill=black] (1.9,0)--(1.7,0)--(1.7,0.2)--(1.8,0.2)--(1.8,0.1)--(1.9,0.1)--cycle;
\draw [fill=black] (1.9,0.4)--(1.7,0.4)--(1.7,0.6)--(1.8,0.6)--(1.8,0.5)--(1.9,0.5)--cycle;

\draw [fill=black] (2.8,0)--(2.6,0)--(2.6,0.1)--(2.7,0.1)--(2.7,0.2)--(2.6,0.2)--(2.6,0.3)--(2.7,0.3)--(2.7,0.4)--(2.8,0.4)--(2.8,0.3)--(2.9,0.3)--(2.9,0.2)--(2.8, 0.2)--cycle;

\draw [fill=black] (2.0,0.8)--(1.8,0.8)--(1.8,0.9)--(1.9,0.9)--(1.9,1)--(1.8,1)--(1.8,1.1)--(1.9,1.1)--(1.9,1.2)--(2.0,1.2)--(2.0,1.1)--(2.1,1.1)--(2.1,1)--(2.0, 1)--cycle;

\draw [fill=black] (2.5,0)--(2.3,0)--(2.3,0.2)--(2.2,0.2)--(2.2,0.3)--(2.3,0.3)--(2.3,0.6)--(2.2,0.6)--(2.2,0.7)--(2.3,0.7)--(2.3,0.8)--(2.4,0.8)--(2.4,0.7)--(2.5, 0.7)--(2.5,0.6)--(2.4,0.6)--(2.4, 0.3)--(2.5,0.3)--(2.5,0.2)--(2.4,0.2)--(2.4,0.1)--(2.5,0.1)--cycle;

 \draw [fill=black](1.6,0.2)rectangle(1.7,0.3);
\draw [fill=black](1.6,0.6)rectangle(1.7,0.7);
\draw [fill=black](1.6,0.8)rectangle(1.7,0.9);
\draw [fill=black](1.6,1)rectangle(1.7,1.1);
\draw [fill=black](1.6,1.4)rectangle(1.7,1.5);

\draw [step=0.1, color=gray!100] (0,0)grid(3.1,1.6);
\end{tikzpicture}
\qquad
\begin{tikzpicture}[scale=1.25]
\draw (1.6,1.9 )node{Rule 90};
\draw [fill=black](0,0)rectangle(0.1 ,0.1);
\draw [fill=black](0.2,0)rectangle(0.3 ,0.1);
\draw [fill=black](0.1,0.1)rectangle(0.2 ,0.2);

\draw [fill=black](0.4,0)rectangle(0.5 ,0.1);
\draw [fill=black](0.6,0)rectangle(0.7 ,0.1);
\draw [fill=black](0.5,0.1)rectangle(0.6 ,0.2);

\draw [fill=black](0.8,0)rectangle(0.9 ,0.1);
\draw [fill=black](1,0)rectangle(1.1 ,0.1);
\draw [fill=black](0.9,0.1)rectangle(1 ,0.2);

\draw [fill=black](1.2,0)rectangle(1.3 ,0.1);
\draw [fill=black](1.4,0)rectangle(1.5 ,0.1);
\draw [fill=black](1.3,0.1)rectangle(1.4 ,0.2);

\draw [fill=black](1.6,0)rectangle(1.7 ,0.1);
\draw [fill=black](1.8,0)rectangle(1.9 ,0.1);
\draw [fill=black](1.7,0.1)rectangle(1.8 ,0.2);

\draw [fill=black](2,0)rectangle(2.1 ,0.1);
\draw [fill=black](2.2,0)rectangle(2.3 ,0.1);
\draw [fill=black](2.1,0.1)rectangle(2.2 ,0.2);

\draw [fill=black](2.4,0)rectangle(2.5 ,0.1);
\draw [fill=black](2.6,0)rectangle(2.7 ,0.1);
\draw [fill=black](2.5,0.1)rectangle(2.6 ,0.2);

\draw [fill=black](2.8,0)rectangle(2.9 ,0.1);
\draw [fill=black](3,0)rectangle(3.1 ,0.1);
\draw [fill=black](2.9,0.1)rectangle(3 ,0.2);

\draw [fill=black](0.2,0.2)rectangle(0.3 ,0.3);
\draw [fill=black](0.4,0.2)rectangle(0.5 ,0.3);
\draw [fill=black](0.3,0.3)rectangle(0.4 ,0.4);

\draw [fill=black](1,0.2)rectangle(1.1 ,0.3);
\draw [fill=black](1.2,0.2)rectangle(1.3 ,0.3);
\draw [fill=black](1.1,0.3)rectangle(1.2 ,0.4);

\draw [fill=black](1.8,0.2)rectangle(1.9 ,0.3);
\draw [fill=black](2,0.2)rectangle(2.1 ,0.3);
\draw [fill=black](1.9,0.3)rectangle(2 ,0.4);

\draw [fill=black](2.6,0.2)rectangle(2.7 ,0.3);
\draw [fill=black](2.8,0.2)rectangle(2.9 ,0.3);
\draw [fill=black](2.7,0.3)rectangle(2.8 ,0.4);

\draw [fill=black](0.4,0.4)rectangle(0.5 ,0.5);
\draw [fill=black](0.6,0.4)rectangle(0.7 ,0.5);
\draw [fill=black](0.5,0.5)rectangle(0.6 ,0.6);

\draw [fill=black](0.8,0.4)rectangle(0.9 ,0.5);
\draw [fill=black](1,0.4)rectangle(1.1 ,0.5);
\draw [fill=black](0.9,0.5)rectangle(1 ,0.6);

\draw [fill=black](2,0.4)rectangle(2.1 ,0.5);
\draw [fill=black](2.2,0.4)rectangle(2.3 ,0.5);
\draw [fill=black](2.1,0.5)rectangle(2.2 ,0.6);

\draw [fill=black](2.4,0.4)rectangle(2.5 ,0.5);
\draw [fill=black](2.6,0.4)rectangle(2.7 ,0.5);
\draw [fill=black](2.5,0.5)rectangle(2.6 ,0.6);

\draw [fill=black](0.6,0.6)rectangle(0.7 ,0.7);
\draw [fill=black](0.8,0.6)rectangle(0.9 ,0.7);
\draw [fill=black](0.7,0.7)rectangle(0.8 ,0.8);

\draw [fill=black](2.2,0.6)rectangle(2.3 ,0.7);
\draw [fill=black](2.4,0.6)rectangle(2.5 ,0.7);
\draw [fill=black](2.3,0.7)rectangle(2.4 ,0.8);

%


\draw [fill=black](1.2,1.2)rectangle(1.3 ,1.3);
\draw [fill=black](1.4,1.2)rectangle(1.5 ,1.3);
\draw [fill=black](1.3,1.3)rectangle(1.4 ,1.4);

\draw [fill=black](1.6,1.2)rectangle(1.7 ,1.3);
\draw [fill=black](1.8,1.2)rectangle(1.9 ,1.3);
\draw [fill=black](1.7,1.3)rectangle(1.8 ,1.4);

\draw [fill=black](0.8,0.8)rectangle(0.9 ,0.9);
\draw [fill=black](1,0.8)rectangle(1.1 ,0.9);
\draw [fill=black](0.9,0.9)rectangle(1 ,1);

\draw [fill=black](1.2,0.8)rectangle(1.3 ,0.9);
\draw [fill=black](1.4,0.8)rectangle(1.5 ,0.9);
\draw [fill=black](1.3,0.9)rectangle(1.4 ,1);

\draw [fill=black](1.6,0.8)rectangle(1.7 ,0.9);
\draw [fill=black](1.8,0.8)rectangle(1.9 ,0.9);
\draw [fill=black](1.7,0.9)rectangle(1.8 ,1);

\draw [fill=black](2,0.8)rectangle(2.1 ,0.9);
\draw [fill=black](2.2,0.8)rectangle(2.3 ,0.9);
\draw [fill=black](2.1,0.9)rectangle(2.2 ,1);

\draw [fill=black](1,1)rectangle(1.1 ,1.1);
\draw [fill=black](1.2,1)rectangle(1.3 ,1.1);
\draw [fill=black](1.1,1.1)rectangle(1.2 ,1.2);

\draw [fill=black](1.8,1)rectangle(1.9 ,1.1);
\draw [fill=black](2,1)rectangle(2.1 ,1.1);
\draw [fill=black](1.9,1.1)rectangle(2 ,1.2);

\draw [fill=black](1.4,1.4)rectangle(1.5 ,1.5);
\draw [fill=black](1.6,1.4)rectangle(1.7,1.5);
\draw [fill=black](1.5,1.5)rectangle(1.6 ,1.6);

 \draw [step=0.1, color=gray!100] (0,0)grid(3.1,1.6);

\end{tikzpicture}
\end{figure}
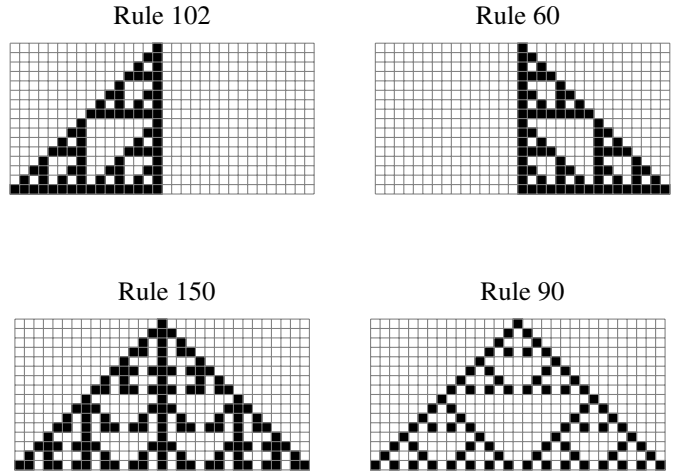

\bigskip
The numbers $01100110$, $00111100$, $10010110$ and $01011010$ are the binary representations of 102, 60, 150 and 90, respectively.
This is the reason why they are called rule 102, rule 60, rule 150 and rule 90.
In Figure~\ref{fig:1}, it is possible to find the mentioned rules using the terminology introduced by Stephen Wolfram~\cite{Wolfram1982}, where a white square represents the digit 0 and a black square represents the digit 1.
 Figure~\ref{fig:2} shows the AC-images generated by these rules after applying 15 iterations to the one-dimensional CA.
 It is possible to see the symmetry between rules 60 and 102.
Notice that, according to Wolfram's terminology, both rules, 60 and 102, are considered for $k=3$, but with one null coefficient.
For example, for rule $102$ the coefficient corresponding to the first component is null, that is, $x_{i}^{t+1}=0\cdot x_{i-1}^t+x_{i}^{t}+x_{i+1}^{t}$.

In Table~\ref{tab:2a} we find an example of a linear, regular, cyclic CA  of length $7$ that uses rule 102. Note that there exists another linear, regular, cyclic CA, with the same length,  that uses rule 60 (see Table~\ref{tab:2b}) and  provides the same sequences. Notice that they appear in reverse order.
Notice that both CAs generate $7$ vertical sequences which are periodic. 
On the other hand, in Table~\ref{tab:ex:90150}, we can find two examples of linear, hybrid, null CA, with length 3, that generate the same sequence (the one in red) as the one in the 102-CA and the 60-CA. In this case, both CAs generate 3 vertical sequences.
From now on, when we say sequence generated by a CA, we mean a vertical sequence. 
\begin{table}[t]
\centering
\caption{Linear, cyclic and regular CA with rules 102 and 60 \label{tab:2}}
\subfloat[\label{tab:2a}]{
\begin{small}\begin{tabular}{|c|c|c|c|c|c|c|}\hline
\textbf{102} & \textbf{102} & \textbf{102} & \textbf{102} & \textbf{102} & \textbf{102} & \textbf{102} \\\hline\hline
 \textcolor{red}{1}&1&1&0&0&1&0\\
\textcolor{red}{0}&0&1&0&1&1&1\\
\textcolor{red}{0}&1&1&1&0&0&1\\
 \textcolor{red}{1}&0&0&1&0&1&1\\
 \textcolor{red}{1}&0&1&1&1&0&0\\
 \textcolor{red}{1}&1&0&0&1&0&1\\
\textcolor{red}{0}&1&0&1&1&1&0\\\hline
\end{tabular}\end{small}
}%
\quad
\subfloat[\label{tab:2b}]{
\begin{small}\begin{tabular}{|c|c|c|c|c|c|c|c|c|c|c|c|c|c|}\hline
\textbf{60} & \textbf{60} & \textbf{60} & \textbf{60} & \textbf{60} & \textbf{60} & \textbf{60} \\\hline\hline
0&1&0&0&1&1& \textcolor{red}{1}\\
1&1&1&0&1&0&\textcolor{red}{0}\\
1&0&0&1&1&1&\textcolor{red}{0}\\
1&1&0&1&0&0& \textcolor{red}{1}\\
0&0&1&1&1&0& \textcolor{red}{1}\\
1&0&1&0&0&1& \textcolor{red}{1}\\
0&1&1&1&0&1&\textcolor{red}{0}\\\hline
\end{tabular}\end{small}
}
\end{table}

\begin{table}[]
    \centering
        \caption{Linear hybrid null CA with rules 150 and 90}
    \label{tab:ex:90150}
    \subfloat[\label{tab:2ca}]{
    \begin{small}
    \begin{tabular}{|c|c|c|}\hline
    90&90&150\\\hline\hline
  \textcolor{red}{1}&0&1\\
  \textcolor{red}{0}&0&1\\
  \textcolor{red}{0}&1&1\\
   \textcolor{red}{1}& 1& 0\\
  \textcolor{red}{1}&1&1\\
  \textcolor{red}{1}&0&0\\
  \textcolor{red}{0}&1&0\\\hline
    \end{tabular}
    \end{small}}\quad 
        \subfloat[\label{tab:2cb}]{
        \begin{small}
    \begin{tabular}{|c|c|c|}\hline
    150&90&90\\\hline\hline
  \textcolor{red}{1}&1&1\\
  \textcolor{red}{0}&0&1\\
  \textcolor{red}{0}&1&0\\
   \textcolor{red}{1}&0 & 1\\
  \textcolor{red}{1}&0&0\\
  \textcolor{red}{1}&1&0\\
  \textcolor{red}{0}& 1&1\\\hline
    \end{tabular}
    \end{small}}
\end{table}

CAs has served as  foundation for  some stream ciphers, thanks to their speed and inherent randomness.
For more information regarding
 implementation in symmetric cryptographic primitives, 
 consult
 \cite{Mariot2024}.
 Additionally, their straightforward hardware implementation and regular structure facilitate the development of efficient software implementations.
The first cryptographic application of CA was published in \cite{Wolfram1985}, where Wolfram used rule 30 in the construction of a stream cipher (broken afterwards by Meier and Stafflebach \cite{Meier1991}).
Other authors have also proposed stream ciphers based on CAs  (see for example \cite{Mihaljevic1998b,Das2013,Jose2014}).

\section{Generators of interleaving sequences  }\label{sec:mismo}
In this section we consider $t$-interleaving sequences where  the corresponding interleaved PN-sequences are shifted versions of the same PN-sequence.
The following result characterises the period and the $LC$ of these sequences. 
More information about these sequences can be found in \cite{Cardell2021a}.

\begin{theorem}
\label{Th:tIL}
Consider a primitive polynomial $p(x)$ of degree $L$. 
If we interleave $2^t$ shifted versions of the   PN-sequence generated by $p(x)$,  then
the resultant $2^t$-interleaving sequence has $LC \leq 2^t\ L$, period $T \leq 2^t  \left(2^L-1\right)$, and it can be generated by $p(x)^{2^t}$.
\end{theorem}
\begin{proof}
    According to \cite[Lemma 3]{Cardell2021a}, 
    if we interleave $2^t$ shifted versions of the same PN-sequence generated by $ p(x)$ of degree $L$, then the resultant $2^t$-interleaving sequence can be generated by $p(x)^{2^t}$.
    This means that the $LC$ 
     at most $2^t\cdot L$.
     Furthermore, if we interleave $2^t$ sequences each with period $2^L-1$, the resulting sequence has period at most $2^t(2^L-1)$. This is because the combined period cannot exceed the product of the number of sequences and their individual periods.
\end{proof}
\begin{remark}
Although the previous theorem can be extended to $t$-interleaving sequences, in this work we restrict our attention to $2^t$-interleaving sequences and thus omit the general case.
\end{remark}

Next, we present two different subsections. 
In Section~\ref{sec:102:mismo} we study the family of 102-CAs that generate these interleaving sequences.
Analogously, in Section~\ref{sec:15090:mismo}, we study the family of 150/90-CAs that generate the same interleaving sequences.

\subsection{The family of 102- CAs}\label{sec:102:mismo}
In this subsection, we investigate the structure of the 102-CAs responsible for generating $t$-interleaving sequences using shifted versions of the same PN-sequence. Notably, all the results derived for 102-CAs can also be formulated with rule 60, as both rules exhibit symmetry (see Section~\ref{sec:prel}).
Notice that due to the complexity and length of the proofs, we present detailed demonstrations only for   $2$-interleaving sequences and $4$-interleaving sequences. These cases already involve intricate arguments and substantial sequence lengths. The general case of $2^t$-interleaving sequences can be deduced by analogy from these smaller cases, as the proofs follow a similar structure. Therefore, we state the general result without repeating the lengthy demonstrations.
\subsubsection{Zech logarithm}
First of all, we need to recall the concept of Zech logarithm which will be useful along the section. 
In this section, we also prove several results that will be useful in the subsequent sections.  
\begin{definition}\label{def:1}
Let $\alpha\in\mathbb{F}_{2^{L}}$ be a primitive element.
The Zech logarithm with basis $\alpha$  is the application $\ensuremath{\mathcal{Z}}_{\alpha}:\mathbb{Z}_{2^{L}-2} \rightarrow \mathbb{Z}_{2^{L}-2}^*\cup \{ \infty\} $, such that each element $t\in\mathbb{Z}_{2^{L}-2}$ corresponds to $\ensuremath{\mathcal{Z}}_{\alpha}(t)$, attaining $1+\alpha^t=\alpha^{\ensuremath{\mathcal{Z}}_{\alpha}(t)}$.
For convenience, we assume that $\alpha^{\infty}=0$.
\end{definition}

\begin{example}\label{ex:zech}
Consider the Galois field $\mathbb{F}_{2^3}$ with $p(x)=1+x+x^3$ as primitive polynomial. 
Let $\alpha$ be a primitive element, root of $p(x)$.
We know that:
$$\begin{array}{|c|c|c|c|c|c|c|c|}\hline
t        & 0 &  1    & 2       & 3       & 4               & 5                  & 6\\\hline
\alpha^t & 1 &\alpha &\alpha^2 &1+\alpha & \alpha+\alpha^2 & 1+ \alpha+\alpha^2 & 1+\alpha^2\\\hline
\end{array}$$
The corresponding Zech logarithms are:
$$\begin{array}{|c|c|c|c|c|c|c|c|}\hline
t                                       & 0 &  1 & 2  & 3 & 4 & 5 & 6\\\hline
\ensuremath{\mathcal{Z}}_{\alpha}(t)    & \infty  &  3 & 6  & 1  & 5   &  4 & 2 \\\hline
\end{array}$$
Since $1+\alpha^0=0$ we consider $\ensuremath{\mathcal{Z}}_{\alpha}(0)=\infty$.  
\end{example}
More properties of the Zech logarithm can be found in \cite{Huber1990}.

 The next result states that the sum of a PN-sequence and a shifted version of itself is again a shifted version of such sequence and the shift can be determined via the Zech logarithm. 
 
 \begin{theorem}\label{lem:2}
Let $p(x)\in\mathbb{F}_2[x]$ be a primitive polynomial of degree $L$ and   $\{a_i\}$ the corresponding PN-sequence.
The sequence $\{a_i+a_{i+k}\}$, with $k\not =0$, is the same PN-sequence $\{a_i\}$  starting at position $D=\ensuremath{\mathcal{Z}}_{\alpha}(k)$ that is $\{a_{i+D}\}= \{a_{i+\ensuremath{\mathcal{Z}}_{\alpha}(k)}\}$.
\end{theorem}

\begin{proof}
Any bit of the PN-sequence $\{a_i\}$ can be computed as
$$a_i=A_0\alpha^i+A_0^2\alpha^{2i}+A_0^4\alpha^{4i}+\cdots+A_0^{2^{L-1}}\alpha^{2^{L-1}i},$$
where $A_0\in\mathbb{F}_{2^{L}}$, and $\alpha\in \mathbb{F}_{2^{L}}$ is a root of $p(x)$ \cite{Lidl1986bk}.
Then, we have that
$$a_i+a_{i+k}=A_0\alpha^i(1+\alpha^k)+A_0^2\alpha^{2i}(1+\alpha^{2k})+A_0^4\alpha^{4i}(1+\alpha^{4k})+\cdots+A_0^{2^{L-1}}\alpha^{2^{L-1}i}(1+\alpha^{k2^{L-1}}).$$
Since $\mathbb{F}_{2^{L}}$ is a field and $\alpha$ is a primitive element, the sum of two elements in the field must be another element in the same field, that is, $1+\alpha^k=\alpha^D$, for some $D\in \{1,2,3, 4, \ldots, 2^{L_2}-2\}$.
Therefore,
$$a_i+a_{i+k}=A_0\alpha^{i+D}+A_0^2\alpha^{2i+2D}+A_0^4\alpha^{4i+4D}+\cdots+A_0^{2^{L_2-1}}\alpha^{2^{L_2-1}i+2^{L_2-1}D}=a_{i+D}.$$
Now, we know that $1+\alpha^k=\alpha^{\ensuremath{\mathcal{Z}}_{\alpha}(k)} $, therefore $D=\ensuremath{\mathcal{Z}}_{\alpha}(k)$ 
and the  theorem is proven.
\end{proof}

\begin{example}\label{ex:1}
Consider  the LFSR  with characteristic polynomial $p_1(x)=1+x+x^3$ and initial state $\{1 1 1\}$.
The corresponding PN-sequence is: $\{a_i\}=\{ 1   1   1   0   0   1   0\}$.
Consider now
$\{a_{i+3}\}=\{ 0   0   1   0   1   1   1\}$, which is the same PN-sequence $\{a_i\}$ starting in position $k=3$. 
The sequence
$\{a_i+a_{i+3}\}=\{1   1   0   0   1   0   1\}$
is the same PN-sequence $\{a_i\}$ starting at position $\ensuremath{\mathcal{Z}}_{\alpha}(3)=1$ (see Example~\ref{ex:zech}).
\end{example}
  
 \begin{corollary}\label{cor:2}
Let $p(x)\in\mathbb{F}_2[x]$ be a primitive polynomial of degree $L$ and   $\{a_i\}$ the corresponding PN-sequence.
If we sum two different shifted versions  $\{a_{i+k_1}\}$   and $\{a_{i+k_2}\}$ of $\{a_i\}$, the resultant sequence $\{a_{i+k_1}+a_{i+k_2}\}$ is a shifted version of $\{a_i\}$  starting at position $k=\ensuremath{\mathcal{Z}}_{\alpha}(k_2-k_1)+k_1$.
\end{corollary}
 
\begin{proof}
 If we denote $\{u_i\}=\{a_{i+k_1}\}$, then $\{u_{i+k_2-k_1}\}=\{a_{i+k_2}\}$.
 Therefore, according to Theorem~\ref{lem:2}, $\{u_{i+k_2-k_1}\}$ is a shifted version of $\{u_i\}$ with a shift $k'=\zech{\alpha}{k_2-k_1}$.
 Now,  $\{u_i\}$ is a shifted version of $\{a_i\}$ with a shift $k_1$.
 Finally, $\{u_{i+k_2-k_1}\}$ is a shifted version of $\{a_i\}$ with a shift   $k=k'+k_1=\zech{\alpha}{k_2-k_1}+k_1.$
\end{proof} 

Now, we are ready to examine the configuration of the CAs responsible for generating interleaving sequences.
First, it is important to note that 102-CAs also generate  individual PN-sequences. 

\begin{theorem}\cite[Theorem 3.3]{Cardell2019bk}
      There exists a regular, cyclic 102-CA of length $\sfrac{\left(2^L-1\right)}{\gcd\left(\zech{\alpha}{1},2^L-1\right)}$,
 that generates the same PN-sequence as that produced by a primitive polynomial $p(x)$ of degree~$L$.
\end{theorem}

\subsubsection{Generating 2-interleaving sequences}\label{sec:102:2}
We start with the study of the generation of 2-interleaving sequences.
 This means that we are interleaving a PN-sequence of period $T=2^L-1$ and linear complexity $L$ with a shifted version of itself.
We consider only the 2-interleaving with maximum period and maximum linear complexity, that is,  $2T$ and $2L$, respectively.
Although sequences with non-maximum linear complexity and period can still be generated by a cellular automata, we will not consider this case, as it is of limited relevance to cryptographic applications.
 From now on, 
 $\alpha$ is the root of the characteristic polynomial of the corresponding PN-sequence.
 
The next result states the structure of the 102-CA that generates such a sequence. 
 
 \begin{theorem}\label{thm:102-2}
 Consider a $2$-interleaving sequence with maximum $LC$ (that is $2L)$.
 There exists a 102-CA (60-CA) of length 
 $$\frac{ 2T}{\gcd(T,D)}$$
 that generates the interleaving sequence in the 0-th column,
where $T$ is the period of the PN-sequence and $D=\ensuremath{\mathcal{Z}}_{\alpha}(1)$.
 \end{theorem}
   \begin{proof}
   Consider the PN-sequence $\{a_i\}$ and the shifted version
   $\{a_{i+k}\}$.
   The corresponding 2-interleaving sequence has the form:
   $$
   \{v_j^{(0)}\}=\{a_0,a_k,a_1,a_{k+1}, a_2, a_{k+2} \ldots \} 
   $$
   Assume we write this sequence in the 0-th column of the 102-CA.
  We   claim that the sequences in the columns with even index, that is, $\{v_j^{(2m)}\}$, with $m=0,1\ldots$ are shifted versions of $ \{v_j^{(0)}\}$.
   
   According to the general form of this 102-CA (see Table~\ref{tab:gen:102CAa} in Appendix 1), the sequence in 
   the first column is 
 $$
   \{v_j^{(1)}\}
   =\{v_0^{(0)}+v_1^{(0)},v_1^{(0)}+v_2^{(0)}, v_2^{(0)}+v_3^{(0)}, \ldots \} 
   =\{a_0+a_k, a_k+a_1,a_1+a_{k+1},a_{k+1}+a_2, \ldots \} 
   $$
  and  the 2-nd column has the form
   $$\{v_j^{(2)}\}
   =\{v_0^{(1)}+v_1^{(1)},v_1^{(1)}+v_2^{(1)}, v_2^{(1)}+v_3^{(1)}, \ldots \} 
   =\{a_0+a_1,a_k+a_{k+1},a_1+a_2, a_{k+1}+ a_{k+2} \ldots \}$$
   that is, $\{v_j^{(2)}\}$ is a 2-interleaving sequence composed of   
   $\{a_i+a_{i+1}\}$ and $\{a_{i+k}+a_{i+k+1}\}$.
   Notice that, according to Theorem~\ref{lem:2},
   $\{a_i+a_{i+1}\}$ is a shifted version of $\{a_i\}$ and  $\{a_{i+k}+a_{i+k+1}\}$ is a shifted version of $\{a_{i+k}\}$ both with a  shift $D=\zech{\alpha}{1}$.
     This means that $\{v_j^{(2)}\}$ is a shifted version of $ \{v_j^{(0)}\}$ with a shift $2 D$.
   
   Using the same argument, we can assure that the sequence in the 4-th  column, $\{v_j^{(4)}\}$, is a shifted version of $\{v_j^{(2)}\}$ with a shift $2D$, that is,  a shifted version of $\{v_j^{(0)}\}$, with a shift $4D$.
   Therefore, the sequence in the $2m$-th column is a shifted version of $\{v_j^{(0)}\}$ with a shift $2mD$.
   
Now, since $2T$ is the period of the 2-interleaving sequence $\{v_j^{(0)}\}$, we know that the column $\{v_j^{(\ell)}\}$, with $\ell= 2 T$ is the same as $ \{v_j^{(0)}\}$, and the shift in this case is $2TD$.
  Therefore, there exists a 102-CA of length $2T$ that generates $\{v_j^{(0)}\}$. 
  However, if $d=\gcd(T,D)\not =1$, the CA can be shorter, since the sequence $ \{v_j^{(0)}\}$ appears in a column prior to the $2T$-th.
  In this case, the column in position
$\frac{ 2T}{\gcd(T,D)}$  is the same as  $ \{v_j^{(0)}\}$
  and the corresponding shift  is $\frac{2TD}{d}$, which is a multiple of~$2T$.
   \end{proof}
   
      \begin{remark}
       If we observe the sequences in  the odd columns, we check that they are also shifted versions of the same sequence with the same shifts. 
        This means that there are shifted versions of two different sequences (one of them is the given 2-interleaving sequence) that appear several times along the CA.
      \end{remark}

   \begin{example}\label{ex:102CA:4}
   
Consider  $p(x)= 1+x^3+x^4$, the PN-sequence
$\{ 1   1   1   1   0   1   0   1   1   0   0   1   0   0   0\}$
and the shifted version
$\{0   1   0   1   1   0   0   1   0   0   0   1   1   1   1\}$.
We can construct a 2-interleaving sequence:
$$\{   1   0   1   1   1   0   1   1   0   1   1   0   0   0   1   1   1   0   0    0   0   0   1   1   0   1   0   1   0   1\}$$
Notice that the 102-CA in Table~\ref{tab:102CA:4} generates this 2-interleaving sequence.
If we consider $\alpha$ a primitive element of $\Fset_{2^4}$, root of $p(x)$, we have  $D=\zech{\alpha}{1}=12$.
Therefore,
$$\frac{2T}{\gcd(T,D)}=\frac{30}{\gcd(15,12)}=10,$$ which is the CA length. 
Notice that the CA length is 10, regardless of the initial state of the PN sequence, as long as the 2-interleaving sequence has maximum linear complexity, that is, $2LC=8$.

Shifted versions of $\{s_j\}$ (the sequence in the 0-th column) appear in columns 2, 4, 6 and 8.
Furthermore, shifted versions of $\{s_j+s_{j+1}\}$ (the sequence in the 1-st column) appear in columns 3, 5, 7 and 9.
The shifts are $2D=24 $, $4D=18$, $6D=12$ and $8D=6$ (modulo 30), respectively (see Table~\ref{tab:102CA:4}). 
   \end{example}
   \begin{table}[t]
       \centering
       \begin{footnotesize}
                 \begin{tabular}{|c|c|c|c|c|c|c|c|c|c|}\hline
       102& 102& 102& 102& 102& 102& 102& 102& 102& 102\\\hline\hline
  \textcolor{red}{1} & 1 & \circledr{0} & \circledr{1}& \circledb{0} & \circledb{0} & \circledg{0} &  \circledg{0} & \circledo{1} & \circledo{0}\\
    \textcolor{red}{0} & 1 & 1 & 1 & 0 & 0 & 0 & 1 & 1 & 1\\
   \textcolor{red}{1}  & 0 & 0 & 1 & 0 & 0 & 1 & 0 & 0 & 1\\
   \textcolor{red}{1}  & 0 & 1 & 1 & 0 & 1 & 1 & 0 & 1 & 0\\
   \textcolor{red}{1}  & 1 & 0 & 1 & 1 & 0 & 1 & 1 & 1 & 1\\
    \textcolor{red}{0} & 1 & 1 & 0 & 1 & 1 & 0 & 0 & 0 & 0\\
\circledo{   \textcolor{red}{1}}  & \circledo{0}& 1 & 1 & 0 & 1 & 0 & 0 & 0 & 0\\
   \textcolor{red}{1}  & 1 & 0 & 1 & 1 & 1 & 0 & 0 & 0 & 1\\
    \textcolor{red}{0} & 1 & 1 & 0 & 0 & 1 & 0 & 0 & 1 & 0\\
   \textcolor{red}{1}  & 0 & 1 & 0 & 1 & 1 & 0 & 1 & 1 & 0\\
   \textcolor{red}{1}  & 1 & 1 & 1 & 0 & 1 & 1 & 0 & 1 & 1\\
   \textcolor{red}{0}  & 0 & 0 & 1 & 1 & 0 & 1 & 1 & 0 & 0\\
   \circledg{ \textcolor{red}{0}} & \circledg{0} & 1 & 0 & 1 & 1 & 0 & 1 & 0 & 0\\
    \textcolor{red}{0} & 1 & 1 & 1 & 0 & 1 & 1 & 1 & 0 & 0\\
   \textcolor{red}{1}  & 0 & 0 & 1 & 1 & 0 & 0 & 1 & 0 & 0\\
   \textcolor{red}{1} & 0 & 1 & 0 & 1 & 0 & 1 & 1 & 0 & 1\\
   \textcolor{red}{1}  & 1 & 1 & 1 & 1 & 1 & 0 & 1 & 1 & 0\\
    \textcolor{red}{0} & 0 & 0 & 0 & 0 & 1 & 1 & 0 & 1 & 1\\
   \circledb{ \textcolor{red}{0}} & \circledb{0} & 0 & 0 & 1 & 0 & 1 & 1 & 0 & 1\\
    \textcolor{red}{0} & 0 & 0 & 1 & 1 & 1 & 0 & 1 & 1 & 1\\
    \textcolor{red}{0} & 0 & 1 & 0 & 0 & 1 & 1 & 0 & 0 & 1\\
    \textcolor{red}{0} & 1 & 1 & 0 & 1 & 0 & 1 & 0 & 1 & 1\\
   \textcolor{red}{1}  & 0 & 1 & 1 & 1 & 1 & 1 & 1 & 0 & 1\\
   \textcolor{red}{1}  & 1 & 0 & 0 & 0 & 0 & 0 & 1 & 1 & 0\\
   \circledr{ \textcolor{red}{0}} & \circledr{1}& 0 & 0 & 0 & 0 & 1 & 0 & 1 & 1\\
   \textcolor{red}{1}  & 1 & 0 & 0 & 0 & 1 & 1 & 1 & 0 & 1\\
    \textcolor{red}{0} & 1 & 0 & 0 & 1 & 0 & 0 & 1 & 1 & 0\\
   \textcolor{red}{1}  & 1 & 0 & 1 & 1 & 0 & 1 & 0 & 1 & 0\\
    \textcolor{red}{0} & 1 & 1 & 0 & 1 & 1 & 1 & 1 & 1 & 1\\
   \textcolor{red}{1}  & 0 & 1 & 1 & 0 & 0 & 0 & 0 & 0 & 1\\\hline
       \end{tabular}
       \end{footnotesize}
       \caption{102-CA that generates the 2-interleaving sequence in Example~\ref{ex:102CA:4}}
       \label{tab:102CA:4}
   \end{table}
The following result  states that all sequences in the 102-CA are 2-interleaving sequences of shifted versions of the PN-sequence. 

 \begin{theorem}\label{thm:102-2b}
Each sequence  in the 102-CA is  a 2-interleaving  sequence  composed of   two shifted versions of $\{a_i\}$ or a 2-interleaving sequence   composed of one shifted version and the zero sequence.  \end{theorem}
   
  \begin{proof}
 
  In Theorem~\ref{thm:102-2}, we stated that several shifted versions of   two sequences appear   along the CA. 
  Obviously, the sequence in the 0-th column is a 
  2-interleaving sequence, since it is constructed in that way, thus the sequences in the columns with even indices are also interleaving sequences. 
  Therefore, it is  enough to prove that the sequence in  the 1-st column is a 2-interleaving sequence  of the form stated in the theorem. 
  
 In this proof we go one step further, we  provide the shifts  corresponding to the PN-sequences  that compose each column.
 In order to do so, we need to consider three different cases, when the shift in the given 2-interleaving sequence is 0, 1 or otherwise. 
 
  \medskip
  
  \pmb{Case $k>1$}

   Consider the PN-sequence $\{a_i\}$ and the shifted version
   $\{a_{i+k}\}$.
   The corresponding 2-interleaving sequence has the form:
   $
   \{v_j^{(0)}\}=\{a_0,a_k,a_1,a_{k+1}, a_2, a_{k+2} \ldots \}
   $
   where
   $$
     v_j^{(0)}=
   \begin{cases}
    a_n & \text{ if } j=2n \text{ for some n},\\
    a_{k+n} & \text{ if } j=2n+1 \text{ for some n},\\
   \end{cases}
   $$
   for $j=0,1,\ldots$.
   If we put the sequence $\{v_j^{(0)}\}$
 in the 0-th column of the 102-CA, then the next sequence of the 102-CA (the first column) has the form:
   $$
  \{v_j^{(1)}\}= \{
   v_0^{(0)}+v_1^{(0)}, v_1^{(0)}+v_2^{(0)}, v_2^{(0)}+v_3^{(0)},\ldots
   \}
   =
      \{
   a_0+a_k, a_k+a_{1}, a_1+a_{k+1},a_{k+1}+a_{2}\ldots 
   \}
   $$
   Notice that this new sequence  $\{v_j^{(1)} \}$ is a 2-interleaving   of the sequences $\{a_i+a_{i+k}\}$ and $\{a_{i+1}+a_{i+k}\}$.
   According to Theorem~\ref{lem:2} and Corollary~\ref{cor:2}, these sequences are shifted versions of $\{a_i\}$  as long as $k>1$ (cases $k=0,1$ are studied  below). 
   The corresponding shifts are $\zech{\alpha}{k}$ and $\zech{\alpha}{k-1}+1$, respectively. 
   Using Corollary~\ref{cor:2} again several times, we check that all sequences of the 102-CA are 2-interleaving sequences of shifted versions of $\{a_i\}$.
In fact, the $j$-th column ($j=0, 1, \ldots$)   is generated by interleaving the two PN-sequences, $\{a_{i+k_1^{(j)}}\}$ and  $\{a_{i+k_2^{(j)}}\}$, which are shifted versions of $\{a_i\}$, with shifts,
   \begin{align*}
     k_1^{(j)}&=k_1^{(j-1)}+ \zech{\alpha}{k_2^{(j-1)}-k_1^{(j-1)}},\\   
     k_2^{(j)}&=k_1^{(j-1)}+1+ \zech{\alpha}{k_2^{(j-1)}-k_1^{(j-1)}-1},\\
   \end{align*}
  where $k_1^{(0)}=0$ and $k_2^{(0)}=k$.  
  
\medskip
  \pmb{ Case $k=0$}
  \medskip
  
  In this case we are interleaving the sequence $\{a_i\}$ with itself.
   According to  Table~\ref{tab:gen:102CAa}, the 1-st column of the 102-CA is an interleaving sequence of 
    $\{a_i+a_{i+k}\}$ and $\{a_{i+1}+a_{i+k}\}$, respectively.
    Since $k=0$, then $\{a_i+a_{i+k}\}$ is the zero sequence and $\{a_{i+1}+a_{i+k}\}=\{a_{i+D}\}$ , with $D=\zech{\alpha}{1}$ (see Theorem~\ref{lem:2}).
    
  Using the same argument several times, we can check that  
  columns with index $2r$ ($r=0, 1, \ldots$) are composed interleaving a shifted version of   $\{a_i\}$, with a shift $k^{(2r)}=r D$, with itself.
  Furthermore, 
 columns with index $2r+1$  ($r=0,1, \ldots$) are composed interleaving the zero sequence and one  shifted version of $\{a_i\}$ (with a shift
  $k^{(2r+1)}=r D$).
    
\medskip
  \pmb{ Case $k=1$}
  \medskip
  
    According to  Table~\ref{tab:gen:102CAa}, the 1-st column of the 102-CA is an interleaving sequence of 
    $\{a_i+a_{i+k}\}$ and $\{a_{i+1}+a_{i+k}\}$, respectively.
    Since $k=1$, then $\{a_i+a_{i+k}\}=\{a_{i+D}\}$ and $\{a_{i+1}+a_{i+k}\}$ is the zero sequence. 
    
    Using the same argument several times, we can state that columns with index $2r$ ($r=0, 1, \ldots$) are composed interleaving two PN-sequences which are shifted versions of $\{a_i\}$, with shifts
   \begin{align*}
     k_1^{(2r)}&=r\cdot \zech{\alpha}{1},\\    k_2^{(2r)}&=r\cdot \zech{\alpha}{1}+1.\\
   \end{align*}
    Besides, columns with index $2r+1$  ($r=0,1, \ldots$) are composed interleaving one  shifted version of $\{a_i\}$ (with shift
   $k_1^{(2r+1)}=(r+1)  \zech{\alpha}{1}$) and   the zero sequence. 
   
Notice that for $r=T=2^L-1$, the sequences start repeating again. 
In case that $T$ is not prime, we can have a  102-CA with   smaller length.
 \end{proof}

      \begin{example}\label{ex:102CA:3}
   Consider the primitive polynomial 
   $p(x)=1+x^2+x^3$, the generated PN-sequence
   $
  \{a_i\}= \{   1   1   1   0   1   0   0
   \}
   $
   and the shifted version
   $\{a_{i+k}\}=\{
         1   1   0   1   0   0   1
   \}$
   where the shift is $k=1$.
   The corresponding 2-interleaving sequence is given by
     $\{s_j\}= \{1   1   1   1   1   0   0   1   1   0   0   0   0   1\}$
     whose characteristic polynomial is
     $p(x)^2=(1+x^2+x^3)^2$ (this means that $LC=2L=6$).
     According to Theorem~\ref{thm:102-2}, there exists a 102-CA of length 
     $$
     \frac{2T}{\gcd(T,D)}=\frac{2\cdot 7}{\gcd(7,5)}=14,
     $$
     where $D=\zech{\alpha}{1}=5$ ($\alpha$ is a primitive element of $\Fset_{2^3}$, root of $p(x)$),
    that generates $\{s_j\}$ in the 0-th column (see Table~\ref{tab:102CA:3}).
    
    Notice that  shifted versions of $\{s_j\}$ appear  in columns 2, 4, 6, 8, 10 and 12.
    Besides, shifted versions of the sequence in the 1-st column, that is $\{s_j+s_{j+1}\}$,  appear in columns 3, 5, 7, 9, 11 and 13.
    The shifts are $2D=10 $, $4D=6$, $6D=2$ , $8D=12$, $10D=8$, $12D=4$ (modulo 14), respectively (see Table~\ref{tab:102CA:3}).

  Observe now, the sequence in the 1-st column, that is,
     $\{0 0 0 0 1 0 1 0 1 0 0 0 1 0 \}$,  is a 2-interleaving sequence composed of 
     $\{a_i+a_{i+k}\}=\{   0   0   1   1   1   0   1\}$ and $\{a_{i+1}+a_{i+k}\}=\{0000000\}$.
     The first sequence  is a shifted version of $\{a_i\}$      with a shift  $\zech{\alpha}{1}=5$  (according to Theorem~\ref{lem:2}).
On the other hand, since $k=1$, the sequence  $\{a_{i+1}+a_{i+k}\}$ is the zero sequence.
Therefore, the sequence  in the 1-st column is the interleaving of
$\{a_{i+5}\}$ and the zero sequence.

Now, the sequence in the 2-nd column is a 2-interleaving sequence  composed of two shifted versions of $\{a_i\}$:
\begin{align*}
   \{a_i+a_{i+1}\}&=\{a_{i+5}\}=\{0   0   1   1   1   0   1\}\\
\{a_{i+k}+a_{i+k+1}\}&=\{a_{i+6}\}=\{0   1   1   1   0   1   0\} 
\end{align*}
This makes total sense, since $\{a_i+a_{i+1}\}=\{a_{i+5}\}$ is obtained
XORing  $\{a_{i+5}\}$  with  the zero sequence.
Furthermore, $\{a_{i+k}+a_{i+k+1}\}$ is a shifted version of \{$a_{i+k}\}$ with a shift $\zech{\alpha}{1}=5$ (according to Theorem~\ref{lem:2}) and  \{$a_{i+k}\}$  is a shifted version of  $\{a_{i}\}$  with  a shift $k=1$. Finally $\{a_{i+k+1}\}$ is a shifted version of  $\{a_{i}\}$  with a shift $k+5=6$.

Following the same idea, we can check that all columns can be generated interleaving  two shifted versions of $\{a_i\} $  or one shifted version of $\{a_i\} $  with the zero sequence.
Given the $j-$th column, $j=0,1,\ldots, 13$, 
in  Table~\ref{tab:102CA:shifts}
we can check the corresponding values for the shifts $k_1^{(j)}$ and $k_2^{(j)}$,  corresponding to the two shifted versions  of $\{a_i\} $, that is $\{a_{i+k_1^{(j)}}\} $ and $\{a_{i+k_2^{(j)}}\} $ . The symbol $-$ correspond to the zero sequence.
   \end{example}
 
\begin{table}
         \centering\begin{footnotesize}
         \begin{tabular}{|c|c|c|c|c|c|c|c|c|c|c|c|c|c|}\hline
         102& 102& 102& 102& 102& 102& 102& 102& 102& 102& 102& 102& 102& 102\\\hline \hline
  \textcolor{red}{1} & 0 & \circledr{0} & \circledr{0} & \circledb{0} & \circledb{1} & \circledg{1} & \circledg{0} & \circledo{0} & \circledo{1} & \circledbl{1} & \circledbl{1} & \circled{1} & \circled{1}  \\
  \textcolor{red}{1} & 0 & 0 & 0 & 1 & 0 & 1 & 0 & 1 & 0 & 0 & 0 & 0 & 0 \\
  \circledg{\textcolor{red}{1}} & \circledg{0} & 0 & 1 & 1 & 1 & 1 & 1 & 1 & 0 & 0 & 0 & 0 & 1\\
  \textcolor{red}{1}& 0 & 1 & 0 & 0 & 0 & 0 & 0 & 1 & 0 & 0 & 0 & 1 & 0\\
  \circled{\textcolor{red}{1}} & \circled{1} & 1 & 0 & 0 & 0 & 0 & 1 & 1 & 0 & 0 & 1 & 1 & 1\\
  \textcolor{red}{0} & 0 & 1 & 0 & 0 & 0 & 1 & 0 & 1 & 0 & 1 & 0 & 0 & 0\\
  \circledb{\textcolor{red}{0}} &  \circledb{1} & 1 & 0 & 0 & 1 & 1 & 1 & 1 & 1 & 1 & 0 & 0 & 0\\
\textcolor{red}{1}& 0& 1 & 0 & 1 & 0 & 0 & 0 & 0 & 0 & 1 & 0 & 0 & 0\\
\circledbl{  \textcolor{red}{1}} & \circledbl{  1} & 1 & 1 & 1 & 0 & 0 & 0 & 0 & 1 & 1 & 0 & 0 & 1\\
  \textcolor{red}{0} & 0 & 0 & 0 & 1 & 0 & 0 & 0 & 1 & 0 & 1 & 0 & 1 & 0\\
\circledr{  \textcolor{red}{0}} &\circledr{ 0} & 0 & 1 & 1 & 0 & 0 & 1 & 1 & 1 & 1 & 1 & 1 & 0\\
  \textcolor{red}{0} & 0 & 1 & 0 & 1 & 0 & 1 & 0 & 0 & 0 & 0 & 0 & 1 & 0\\
    \circledo{\textcolor{red}{0}} &   \circledo{1} & 1 & 1 & 1 & 1 & 1 & 0 & 0 & 0 & 0 & 1 & 1 & 0\\
  \textcolor{red}{1} & 0 & 0 & 0 & 0 & 0 & 1 & 0 & 0 & 0 & 1 & 0 & 1 & 0\\\hline
         \end{tabular}
         \end{footnotesize}
         \caption{102-CA that generates the 2-interleaving sequence of Example~\ref{ex:102CA:3}}
         \label{tab:102CA:3}
     \end{table}
     
    \begin{table}[]
        \centering
        \begin{tabular}{|c|cccccccccccccc|}\cline{2-15}
        \multicolumn{1}{c|}{}&\multicolumn{14}{|c|}{$j$}\\\cline{2-15}
      \multicolumn{1}{c|}{}  &0 & 1 & 2 & 3 & 4 & 5 &6&7&8&9&10&11&12&13\\\hline
$k_1^{(j)}$   & 0 & 5 & 5 & 3 & 3 & 1 & 1 & 6 & 6 & 4 & 4 & 2 & 2 & 0\\
$k_2^{(j)}$    &1 & - & 6  &- & 4  &- & 2 & - & 0 & - & 5 & - & 3 & -\\\hline
        \end{tabular}
        \caption{Shifts corresponding to the interleaving PN-sequences in the  $j$-th column  of the 102-CA in Table~\ref{tab:102CA:3}}
        \label{tab:102CA:shifts}
    \end{table} 






  \subsubsection{Generating 4-interleaving sequences}
  
  Now, we discuss the case of interleaving  4 PN-sequences. 
  We present a  brief study, since the results are very similar to those of Section~\ref{sec:102:2}.
  We do not consider the case of interleaving 3 sequences, as the resulting behaviour does not exhibit a clear or consistent pattern. Therefore, we restrict our study to powers of two.
 \begin{theorem}
 If the $4$-interleaving sequence has maximum $LC$ (that is $4L)$, then
 there exists a 102-CA (60-CA) of length 
 $$\frac{ 4T}{\gcd(T,D)},$$
where $T$ is the period of the PN-sequence sequence and $D=\ensuremath{\mathcal{Z}}_{\alpha}(1)$,
which generates the interleaving sequence in the 0-th column. 
 \end{theorem}
  
\begin{proof}  
     Consider the PN-sequence $\{a_i\}$ and the shifted versions
   $\{a_{i+k_1}\}$,   $\{a_{i+k_2}\}$ and   $\{a_{i+k_3}\}$.
   The corresponding 4-interleaving sequence has the form:
   $$
   \{v_j^{(0)}\}=\{a_0,a_{k_1},a_{k_2}, a_{k_3}, a_1,a_{k_1+1}, a_{k_2+1}, a_{k_3+1}, \ldots \}$$
   We can write this sequence in the 0-th column of the 102-CA.
   We prove that the sequences in the columns of the form $\{v_j^{(4m)}\}$, with $m=0,1\ldots$ are shifted versions of $ \{v_j^{(0)}\}$.
   
   According to the general form of this 102-CA (see Table~\ref{tab:gen:102CAb} em Appendix 1), the column in the 4-th column  has the form
   $$\{v_j^{(4)}\}=\{a_0+a_1,a_k+a_{k+1},a_1+a_2, a_{k+1}+ a_{k+2} \ldots \},$$
   that is, it is a 4-interleaving sequence  composed by the sequences
   $\{a_i+a_{i+1}\}$, $\{a_{i+k_1}+a_{i+k_1+1}\}$, $\{a_{i+k_2}+a_{i+k_2+1}\}$ and  $\{a_{i+k_3}+a_{i+k_3+1}\}$.
   Notice that, according to Theorem~\ref{lem:2},
   $\{a_i+a_{i+1}\}=\{a_{i+D}\}$ is a shifted version of $\{a_i\}$,  and  $\{a_{i+k_j}+a_{i+k_j+1}\}=\{a_{i+k_j+D}\}$ is a shifted version of $\{a_{i+k_j}\}$,  for $j=1,2,3$, where  $D=\zech{\alpha}{1}$.
      This means that $\{v_j^{(4)}\}$ is a shifted version of $ \{v_j^{(0)}\}$ with a shift $4 D$.
   
   Using the same argument, we can ensure that the sequence in the 8-th column, $\{v_j^{(8)}\}$ is a shifted version of $\{v_j^{(4)}\}$ with a shift $4 D$, that is,  a shifted version of $\{v_j^{(0)}\}$, with a shift $8D$.
   Therefore, the sequence in the $4m$-th column is a shifted version of $\{v_j^{(0)}\}$ with a shift $4mD$.
   
    If $4T$ is the period of the 4-interleaving sequence $\{v_j^{(0)}\}$, we know that the column $\{v_j^{(\ell)}\}$, with $\ell= 4 T$ is the same as $ \{v_j^{(0)}\}$, since the shift in this case is $4TD$.
  Therefore, there exists a   102-CA of length $4T$, that generates such a sequence. 
  However, if $d=\gcd(T,D)\not =1$, the CA can be shorter.
  In this case, the 102-CA has length
$\frac{ 4T}{\gcd(T,D)}$  
  and the shift is $\frac{4TD}{d}$, which is a multiple of $4T$.
  \end{proof}

  \begin{remark}
       If we observe the other sequences in the 102-CA, it is possible to check that 
        there are other 3 sequences that appear along the CA with different shifts. 
            This means that there are four different sequences (including the given 4-interleaving sequence)  and shifted versions of these 4 sequences that appear along the CA, always with the same shift. 
      \end{remark}
      
   \begin{theorem}All the sequences in the 102-CA are 4-interleaving sequences composed of shifted versions of $\{a_i\}$ or/and the zero column.
 \end{theorem}
We skip the proof of this theorem, since it is analogous to the one of Theorem~\ref{thm:102-2b}.
  \begin{example}\label{ex:102CA:3x4}
  Consider again the primitive polynomial $p(x)=1+x^2+x^3$, the PN-sequence
 $\{a_i\}= \{   1   0   0   1   1   1   0\}$
  and the shifted versions:
$
  \{a_{i+5}\}=\{   1   0   1   0   0   1   1\},
  \{a_{i+4}\}=\{   1   1   0   1   0   0   1\},
  \{a_{i+1}\}=\{   0   0   1   1   1   0   1\}
$
 The corresponding 4-interleaving sequence  has the form:
 $$
 \{s_j\}=\{
    1   1   1   0   0   0   1   0   0   1   0   1   1   0   1   1   1   0   0   1   1   1   0   0   0   1   1   1
 \}
 $$
 There exists a 102-CA of length
  $$\frac{ 4T}{\gcd(T,D)}=\frac{4\cdot
  7}{\gcd(7,5)}=28$$
 that generates the sequence $\{s_j\}$ in the 0-th column (see Table~\ref{tab:102CA:3x4}). 
 Notice that  other 6 different shifted versions of each one of the fist 4 sequences (including the given 4-interleaving sequence)  appear along the CA.
 The shifts of the shifted versions are $4D=20$, $8D=12$, $12D=4$, $16D=24 $, $20D=16$ and $24D=8$ $(\text{mod } 28)$, respectively (see Table~\ref{tab:102CA:3x4}).
 
  \end{example}
  \begin{table}[]
      \centering
      \begin{footnotesize}
   \begin{tabular}{|p{1pt}|p{1pt}|p{1pt}|p{1pt}|p{1pt}|p{1pt}|p{1pt}|p{1pt}|p{1pt}|p{1pt}|p{1pt}|p{1pt}|p{1pt}|p{1pt}|p{1pt}|p{1pt}|p{1pt}|p{1pt}|p{1pt}|p{1pt}|p{1pt}|p{1pt}|p{1pt}|p{1pt}|p{1pt}|p{1pt}|p{1pt}|p{1pt}|}\hline
   \multicolumn{28}{|c|}{102}\\\hline\hline
  1&0&0&1&\pmb{\textcolor{red}{1}}&
 \pmb{\textcolor{red}{0}}&\pmb{\textcolor{red}{1}}&\pmb{\textcolor{red}{0}}&\pmb{\textcolor{blue}{1}}&\pmb{\textcolor{blue}{1}}&\pmb{\textcolor{blue}{0}}&\pmb{\textcolor{blue}{1}}&\pmb{\textcolor{green}{0}}&\pmb{\textcolor{green}{0}}&\pmb{\textcolor{green}{1}}&\pmb{\textcolor{green}{1}}&\pmb{\textcolor{yellow}{0}}&\pmb{\textcolor{yellow}{1}}&\pmb{\textcolor{yellow}{1}}&\pmb{\textcolor{yellow}{1}}&\pmb{\textcolor{purple!50}{1}}&\pmb{\textcolor{purple!50}{1}}&\pmb{\textcolor{purple!50}{1}}&\pmb{\textcolor{purple!50}{0}}&\pmb{\textcolor{orange}{0}}&\pmb{\textcolor{orange}{1}}&\pmb{\textcolor{orange}{0}}&\pmb{\textcolor{orange}{0}}\\
  1&0&1&0&1&1&1&1&0&1&1&1&0&1&0&1&1&0&0&0&0&0&1&0&1&1&0&1\\
  1&1&1&1&0&0&0&1&1&0&0&1&1&1&1&0&1&0&0&0&0&1&1&1&0&1&1&0\\
  0&0&0&1&0&0&1&0&1&0&1&0&0&0&1&1&1&0&0&0&1&0&0&1&1&0&1&1\\
  \pmb{\textcolor{green}{0}}&\pmb{\textcolor{green}{0}}&\pmb{\textcolor{green}{1}}&\pmb{\textcolor{green}{1}}&0&1&1&1&1&1&1&0&0&1&0&0&1&0&0&1&1&0&1&0&1&1&0&1\\
  0&1&0&1&1&0&0&0&0&0&1&0&1&1&0&1&1&0&1&0&1&1&1&1&0&1&1&1\\
  1&1&1&0&1&0&0&0&0&1&1&1&0&1&1&0&1&1&1&1&0&0&0&1&1&0&0&1\\
  0&0&1&1&1&0&0&0&1&0&0&1&1&0&1&1&0&0&0&1&0&0&1&0&1&0&1&0\\
  \pmb{\textcolor{orange}{0}}&\pmb{\textcolor{orange}{1}}&\pmb{\textcolor{orange}{0}}&\pmb{\textcolor{orange}{0}}&1&0&0&1&1&0&1&0&1&1&0&1&0&0&1&1&0&1&1&1&1&1&1&0\\
  1&1&0&1&1&0&1&0&1&1&1&1&0&1&1&1&0&1&0&1&1&0&0&0&0&0&1&0\\
  0&1&1&0&1&1&1&1&0&0&0&1&1&0&0&1&1&1&1&0&1&0&0&0&0&1&1&1\\
  1&0&1&1&0&0&0&1&0&0&1&0&1&0&1&0&0&0&1&1&1&0&0&0&1&0&0&1\\
  \pmb{\textcolor{blue}{1}}&\pmb{\textcolor{blue}{1}}&\pmb{\textcolor{blue}{0}}&\pmb{\textcolor{blue}{1}}&0&0&1&1&0&1&1&1&1&1&1&0&0&1&0&0&1&0&0&1&1&0&1&0\\
  0&1&1&1&0&1&0&1&1&0&0&0&0&0&1&0&1&1&0&1&1&0&1&0&1&1&1&1\\
  1&0&0&1&1&1&1&0&1&0&0&0&0&1&1&1&0&1&1&0&1&1&1&1&0&0&0&1\\
1&0&1&0&0&0&1&1&1&0&0&0&1&0&0&1&1&0&1&1&0&0&0&1&0&0&1&0\\
  \pmb{\textcolor{purple!50}{1}}&\pmb{\textcolor{purple!50}{1}}&\pmb{\textcolor{purple!50}{1}}&\pmb{\textcolor{purple!50}{0}}&0&1&0&0&1&0&0&1&1&0&1&0&1&1&0&1&0&0&1&1&0&1&1&1\\
  0&0&1&0&1&1&0&1&1&0&1&0&1&1&1&1&0&1&1&1&0&1&0&1&1&0&0&0\\
  0&1&1&1&0&1&1&0&1&1&1&1&0&0&0&1&1&0&0&1&1&1&1&0&1&0&0&0\\
  1&0&0&1&1&0&1&1&0&0&0&1&0&0&1&0&1&0&1&0&0&0&1&1&1&0&0&0\\
   \pmb{\textcolor{red}{1}}&\pmb{\textcolor{red}{0}}&\pmb{\textcolor{red}{1}}&\pmb{\textcolor{red}{0}}&1&1&0&1&0&0&1&1&0&1&1&1&1&1&1&0&0&1&0&0&1&0&0&1\\
  1&1&1&1&0&1&1&1&0&1&0&1&1&0&0&0&0&0&1&0&1&1&0&1&1&0&1&0\\
  0&0&0&1&1&0&0&1&1&1&1&0&1&0&0&0&0&1&1&1&0&1&1&0&1&1&1&1\\
  0&0&1&0&1&0&1&0&0&0&1&1&1&0&0&0&1&0&0&1&1&0&1&1&0&0&0&1\\
\pmb{\textcolor{yellow}{0}}&\pmb{\textcolor{yellow}{1}}&\pmb{\textcolor{yellow}{1}}&\pmb{\textcolor{yellow}{1}}&1&1&1&0&0&1&0&0&1&0&0&1&1&0&1&0&1&1&0&1&0&0&1&1\\
  1&0&0&0&0&0&1&0&1&1&0&1&1&0&1&0&1&1&1&1&0&1&1&1&0&1&0&1\\
  1&0&0&0&0&1&1&1&0&1&1&0&1&1&1&1&0&0&0&1&1&0&0&1&1&1&1&0\\
  1&0&0&0&1&0&0&1&1&0&1&1&0&0&0&1&0&0&1&0&1&0&1&0&0&0&1&1\\\hline
     \end{tabular}
      \end{footnotesize}
      \caption{102-CA that generates the 4-interleaving sequence in Example~\ref{ex:102CA:3x4}}
      \label{tab:102CA:3x4}
  \end{table}
  \subsubsection{Generating $2^t$-interleaving sequences}
   In this section, we consider the resultant sequence of interleaving $2^t$ versions of the same PN-sequence.
   The next results can be derived using the same ideas applied in the preceding sections.

 \begin{theorem}
 Consider a $2^t$-interleaving sequence with maximum $LC$ (that is $2^tL)$.
 There exists a 102-CA (60-CA) of length 
 $$\frac{ 2^tT}{\gcd(T,D)},$$
where $T$ is the period of the PN-sequence sequence and $D=\ensuremath{\mathcal{Z}}_{\alpha}(1)$,
which generates the $2^t$-interleaving sequence in the 0-th column. 
 \end{theorem}
 
 In this case, there are $2^t$ different sequences that appear several times along the CA with different shifts. 
 That is, shifted versions of these sequences appear along the CA with shifts
 $k\ 2^t \ D$, for $k=1, 2, \ldots$.
 Furthermore, all sequences in the 102-CA are $2^t$-interleaving sequences of shifted versions of the PN-sequence and/or the zero sequence.

It is natural to wonder what happens when we interleave $t$ sequences where $t$ is not a power of two.
In that case, there does not seem to be a fixed value for the CA length. There is always a 102-CA that generates such sequences, but the length seems to be quite random (see Table~\ref{tab:CA:t}).

 \begin{table}
 $$
 \begin{array}{|c|c|c|c|}\hline
\text{\backslashbox{$t$}{$L$}}&3 & 4 &5\\\hline
3                  & 3^2\cdot 7 &  3^4\cdot 5^2  &  11\cdot31 \\
5                  & 7^2\cdot 5^2 &  3^2\cdot 5^2 &5^2\cdot 31^2\\
6 & 2\cdot3^2\cdot7 & 2^2\cdot 3^4\cdot5^2 &2\cdot11\cdot31\\
7& 7^4 & 3^2\cdot 5 \cdot 7 \cdot 13  &7\cdot 31\cdot 151
\\ \hline
 \end{array}
   $$
   \caption{Upper bound for the length of the 102-CA that generates a $t$-interleaving sequence \label{tab:CA:t}}
     \end{table}

   \subsection{The family of 150/90-CAs} \label{sec:15090:mismo}
   In this section we study the family of 150/90-CAs that generate $t$- interleaving sequences. 
   
\subsubsection{Generating PN-sequences}
Given a PN-sequence produced by a primitive polynomial of degree $L$,
the Cattell-Muzio algorithm \cite{Cattell1996}  provides two hybrid, null 150/90-CAs of length $L$ that generate such a PN-sequence.
However, the authors did not mention  that all the vertical sequences obtained in these CAs are shifted versions of the same PN-sequence.
This happens because we consistently perform XOR operations on the PN-sequence with itself, as was the case with 102-CA.
If the rule acting in the 0-th column is 90, the next sequence is just the same, shifted one position (see Table~\ref{tab:90}).
On the other hand, if it is 150, then rule 150 acts like rule 102, since the CAs we are considering in this section are null  (see Table~\ref{tab:150}).
The next sequence is then $\{a_i+a_{i+1}\}=\{a_{i+k}\}$, with $k=\zech{\alpha}{1}$.

\begin{table}
    \centering
    \subfloat[\label{tab:90}]
    {
  $ \begin{array}{|c|c|}\hline
    90&-\\\hline\hline
      a_0   &  a_1\\
      a_1   & a_2\\
      a_2 & a_3\\
      \vdots &\vdots\\\hline
    \end{array}$
    }
    \quad
    \subfloat[\label{tab:150}]{
    $\begin{array}{|c|c|} \hline
        150&-\\\hline\hline
      a_0   &  a_0+a_1\\
      a_1   & a_1+a_2\\
      a_2 & a_2+a_3\\
      \vdots &\vdots\\\hline
    \end{array}$}

    \subfloat[\label{tab:90b}]{
    $\begin{array}{|c|c|c|} \hline
        -&90&-\\\hline\hline
      a_{k_1}    &   a_{k_2} &     a_{k_1} +a_{k_2+1}   \\
      a_{k_1+1}  & a_{k_2+1} &   a_{k_1+1} +a_{k_2+2} \\
      a_{k_1+2}  & a_{k_2+2} &   a_{k_1+2} +a_{k_2+3} \\
      \vdots &\vdots&\vdots\\\hline
    \end{array}$}
    \quad
    \subfloat[\label{tab:150b}]{
    $\begin{array}{|c|c|c|} \hline
        -&150&-\\\hline\hline
      a_{k_1}    &   a_{k_2} &     a_{k_1}+a_{k_2} +a_{k_2+1}   \\
      a_{k_1+1}  & a_{k_2+1} &   a_{k_1+1} +a_{k_2+1}+a_{k_2+2} \\
      a_{k_1+2}  & a_{k_2+2} &   a_{k_1+2}+a_{k_2+2} +a_{k_2+3} \\
      \vdots &\vdots&\vdots\\\hline
    \end{array}$}
    \caption{Behaviour of rules 90 and 150 in the CA}
    \label{tab:3}
\end{table}

Now, if rule 90  acts in the middle of the CA, we also have that the next sequence is a PN-sequence  (see Table~\ref{tab:90b}).
In fact, this sequence is given by
$\{a_{i+k_1}+a_{i+k_2}\}=\{a_{i+k}\}$ where $k=\zech{\alpha}{k_2-k_1}+k_1$ (see Corollary \ref{cor:2}).
On the other hand, if rule 150  acts in the middle of the CA, then the next sequence is a PN-sequence (see Table~\ref{tab:150b}).
In fact, this sequence is given by
$\{a_{i+k_1}+a_{i+k_2}+a_{i+k_2+1}\}=\{a_{i+k'}\}$ where
\begin{align*}
    k'=\zech{\alpha}{k_2+1-k}+k,
\end{align*} 
where $k=\zech{\alpha}{k_2-k_1}+k_1$.
Next example illustrates the previous ideas. 

\begin{example}
Consider the primitive polynomial $p(x)=1+x^2+x^3$. According to the   Cattell-Muzio algorithm, the corresponding CAs that generate the PN-sequences generated by $p(x)$ are given by
$$
\begin{array}{|c|c|c|}\hline
0&0&1\\\hline
\end{array}
\qquad
\text{ and }
\qquad
\begin{array}{|c|c|c|}\hline
1&0&0\\\hline
\end{array}
$$
where 1 and 0 represent rules 150 and 90, respectively (notation used by Cattell-Muzio). 
Consider the initial state $\{100\}$,
the corresponding  PN-sequence $\{1001110\}$ can be generated by the two 150/90-CAs represented in Table~\ref{tab:ex:90150}.
Notice that in both CAs, all the sequences are shifted versions of the red one. 
\end{example}

\subsubsection{Generating $2^t$-interleaving sequences }

In \cite{Fuster2007c}, the authors proposed an algorithm to determine two 150/90-CAs that generate  the shrunken sequence. Since the shrunken sequence is an interleaving sequence (see \cite{Cardell2020b} for more details), it is natural to assume that a similar process may work for arbitrary interleaving sequences.

Assume that we have a $2^t$-interleaving sequence with maximum $LC$ and  $p(x)$ as the primitive polynomial of degree $L$ that generates the PN-sequences. 
Next algorithm shows how to obtain two 105/90-CA that generates this interleaving sequence. 

\begin{enumerate}
    \item Apply the Cattel-Muzio algorithm \cite{Cattell1996} to determine two linear 150/90-CA that generate the PN-sequences with characteristic polynomial $p(x)$.
    \item For each CA proceed:
    \begin{itemize}
        \item[3.1] Complement its least significant bit. The resultant string is denoted by $S_i$, $i=1,2$.
        \item[3.2] Compute the mirror image of $S_i$, denoted by $S_i^*$, and concatenate both strings: $$S_i'=[S_i,S_i^*]$$
        \item[3.3] Apply 3.1 and 3.2 to each $S_ i'$ recursively $t$ times. 
     \end{itemize}
\end{enumerate}

As a consequence, we can introduce the following result.

\begin{theorem}
Given a $2^t$-interleaving sequence composed of shifted versions of a PN-sequence with $LC=L$, there exists a 150/90-CA of length $2^t L$ that generates such a sequence. 
\end{theorem}

Next example illustrates the process explained above.

   \begin{example} \label{ex:5:2:1}
Consider the primitive polynomial $p(x)=1+x^2+x^5$, the PN-sequence 
$$
\{
   1   1   1   1   1   0   0   0   1   1   0   1   1   1   0   1   0   1   0   0   0   0   1   0   0   1   0   1   1   0   0
\}
$$
and  a shifted version 
$$
\{
     1   0   0   0   0   1   0   0   1   0   1   1   0   0   1   1   1   1   1   0   0   0   1   1   0   1   1   1   0   1   0
\}.
$$
If we interleave these two sequences, we obtain a $2$-interleaving sequence of the form
$$
\{
       1   1   1   0   1   0   1   0   1   0   0   1   0   0   0   0   1   1   1   0   0   1   1   1   1   0   1   0   0   1   1   1   0   1   1   1   0   1   0   0
   0   0   0   0   1   1   0   1   0   0   1   1   0   1   1   1   1   0   0   1   0   0
\}
$$
with period $T=62$ and $LC=10$ (characteristic polynomial $p(x)^2=(1+x^2+x^5)^2$).

\begin{table}[]
    \centering
 $$
\begin{array}{|c|c|c|}\hline
p(x) & \multicolumn{2}{c|}{150/90-\text{CA} }\\\hline
 1+x^2+x^5   
 & 
 \begin{array}{|c|c|c|c|c|}\hline
  1&1&1&1&0\\ \hline
 \end{array} 
 & 
 \begin{array}{|c|c|c|c|c|}\hline
  0&1&1&1&1\\ \hline 
 \end{array}\\ 
1+x^3+x^5    
& 
 \begin{array}{|c|c|c|c|c|}\hline
0 &1 &1 &0 &0 \\ \hline
 \end{array}
& 
 \begin{array}{|c|c|c|c|c|}\hline
0& 0& 1& 1 &0 \\ \hline 
 \end{array}\\ 
1+x+x^2+x^4+x^5
&
 \begin{array}{|c|c|c|c|c|}\hline
1&0&0&0&0\\ \hline
 \end{array}
&
 \begin{array}{|c|c|c|c|c|}\hline
0&0&0&0&1 \\ \hline
\end{array}\\ 
1+x+x^3+x^4+x^5 
&
 \begin{array}{|c|c|c|c|c|}\hline
1&1&1&0&0\\ \hline
 \end{array}
&
 \begin{array}{|c|c|c|c|c|}\hline
0&0&1&1&1\\ \hline
 \end{array}\\ 
1+x+x^2+x^3+x^5x^5   
& 
 \begin{array}{|c|c|c|c|c|}\hline
1&1&0&0&0 \\ \hline
 \end{array}
&  
 \begin{array}{|c|c|c|c|c|}\hline
0&0&0&1&1\\ \hline
 \end{array}\\ 
1+x^2+x^3+x^4+x^5
&  
 \begin{array}{|c|c|c|c|c|}\hline
1&0&0&1&1\\ \hline
 \end{array}
&   
 \begin{array}{|c|c|c|c|c|}\hline
1&1&0&0&1\\ \hline
 \end{array}
\\ \hline
\end{array}
$$
    \caption{150/90-CAs that generate the PN-sequences of $p(x)$ (1 and 0 represent rules 150 and 90, respectively)}
    \label{tab:CA:grad5}
\end{table}

According to   Cattell\&Muzio   \cite{Cattell1996}, every PN-sequence produced by a primitive polynomial of degree $L$ can be generated by a 150/90-CA of length $L$.
For instance, Table~\ref{tab:CA:grad5} shows the 150/90-CA that generate the PN-sequences of every primitive polynomial of degree 5. 
In particular, the PN-sequences produced by $p(x)$ can be generated by the CAs
$$
\begin{tabular}{r|c|c|c|c|c|}\cline{2-6}
\multicolumn{1}{c|}{\phantom{$s_1:$}}  &0&1&1&1&1\\\cline{2-6}
\end{tabular}
$$
$$
\begin{tabular}{r|c|c|c|c|c|}\cline{2-6}
\multicolumn{1}{c|}{\phantom{$s_2:$}}  &1&1&1&1&0\\ \cline{2-6}
\end{tabular}
$$
where 1 and 0 represent rules 150 and 90, respectively.
Now, we complement their least significant bit:

$$
\begin{tabular}{r|c|c|c|c|c|}\cline{2-6}
\multicolumn{1}{c|}{$S_1:$}  &0&1&1&1&\textcolor{red}{0}\\\cline{2-6}
\end{tabular}
$$
$$
\begin{tabular}{r|c|c|c|c|c|}\cline{2-6}
\multicolumn{1}{c|}{$S_2:$}  &1&1&1&1&\textcolor{red}{1}\\ \cline{2-6}
\end{tabular}
$$
Since we are interleaving 2 PN-sequences, we need to do the mirror process just once.
Therefore, the CAs that generate the given 2-interleaving sequence are:
$$
\begin{tabular}{r|c|c|c|c|c||c|c|c|c|c|}\cline{2-11}
\multicolumn{1}{c|}{$[S_1,S_1^*]:$} &0&1&1&1&0&0&1&1&1&0\\\cline{2-11}
\end{tabular}
$$
$$
\begin{tabular}{r|c|c|c|c|c||c|c|c|c|c|}\cline{2-11}
\multicolumn{1}{c|}{$[S_2,S_2^*]:$} &1&1&1&1&1&1&1&1&1&1\\\cline{2-11}
\end{tabular}
$$
that is:
$$
\begin{tabular}{|c|c|c|c|c|c|c|c|c|c|}\hline
  90&150&150&150&90&90&150&150&150&90\\\hline
\end{tabular}
$$
$$
\begin{tabular}{|c|c|c|c|c|c|c|c|c|c|}\hline
150&150&150&150&150&150&150&150&150&150\\\hline
\end{tabular}
$$

In Table~\ref{tab:CA:52a}, we can find both 90/150-CA. Notice that the corresponding 2-interleaving sequences is generated in the 0-th column (in red) in both cases. 
 \end{example}
 

\begin{table}[]
    \centering
        \caption{150/90-CAs that generate  the 2-interleaving sequence given in Example~\ref{ex:5:2:1}}\label{tab:CA:52a}
    \subfloat[\label{tab:CA:52aa}]{
        \begin{tiny}
              \begin{tabular}{|c|c|c|c|c|c|c|c|c|c|}\hline
   90&150&150&150&90&90&150&150&150&90\\\hline
\textcolor{red}{1} & 1 & 1 & 0 & 0 & 0 & 0 & 1 & 0 & 1\\
\textcolor{red}{1} & 1 & 0 & 1 & 0 & 0 & 1 & 1 & 0 & 0\\
\textcolor{red}{1} & 0 & 0 & 1 & 1 & 1 & 0 & 0 & 1 & 0\\
  \textcolor{red}{0} & 1 & 1 & 0 & 0 & 1 & 1 & 1 & 1 & 1\\
\textcolor{red}{1} & 0 & 0 & 1 & 1 & 1 & 1 & 1 & 1 & 1\\
  \textcolor{red}{0} & 1 & 1 & 0 & 0 & 0 & 1 & 1 & 1 & 1\\
\textcolor{red}{1} & 0 & 0 & 1 & 0 & 1 & 0 & 1 & 1 & 1\\
  \textcolor{red}{0} & 1 & 1 & 1 & 0 & 0 & 0 & 0 & 1 & 1\\
\textcolor{red}{1} & 0 & 1 & 0 & 1 & 0 & 0 & 1 & 0 & 1\\
  \textcolor{red}{0} & 0 & 1 & 0 & 0 & 1 & 1 & 1 & 0 & 0\\
  \textcolor{red}{0} & 1 & 1 & 1 & 1 & 1 & 1 & 0 & 1 & 0\\
\textcolor{red}{1} & 0 & 1 & 1 & 0 & 0 & 0 & 0 & 1 & 1\\
  \textcolor{red}{0} & 0 & 0 & 0 & 1 & 0 & 0 & 1 & 0 & 1\\
  \textcolor{red}{0} & 0 & 0 & 1 & 0 & 1 & 1 & 1 & 0 & 0\\
  \textcolor{red}{0} & 0 & 1 & 1 & 0 & 1 & 1 & 0 & 1 & 0\\
  \textcolor{red}{0} & 1 & 0 & 0 & 0 & 1 & 0 & 0 & 1 & 1\\
\textcolor{red}{1} & 1 & 1 & 0 & 1 & 0 & 1 & 1 & 0 & 1\\
 \textcolor{red}{1} & 1 & 0 & 0 & 0 & 0 & 0 & 0 & 0 & 0\\
\textcolor{red}{1} & 0 & 1 & 0 & 0 & 0 & 0 & 0 & 0 & 0\\
  \textcolor{red}{0} & 0 & 1 & 1 & 0 & 0 & 0 & 0 & 0 & 0\\
  \textcolor{red}{0} & 1 & 0 & 0 & 1 & 0 & 0 & 0 & 0 & 0\\
\textcolor{red}{1} & 1 & 1 & 1 & 0 & 1 & 0 & 0 & 0 & 0\\
\textcolor{red}{1} & 1 & 1 & 0 & 0 & 0 & 1 & 0 & 0 & 0\\
\textcolor{red}{1} & 1 & 0 & 1 & 0 & 1 & 1 & 1 & 0 & 0\\
\textcolor{red}{1} & 0 & 0 & 1 & 0 & 1 & 1 & 0 & 1 & 0\\
  \textcolor{red}{0} & 1 & 1 & 1 & 0 & 1 & 0 & 0 & 1 & 1\\
\textcolor{red}{1} & 0 & 1 & 0 & 0 & 0 & 1 & 1 & 0 & 1\\
  \textcolor{red}{0} & 0 & 1 & 1 & 0 & 1 & 0 & 0 & 0 & 0\\
  \textcolor{red}{0} & 1 & 0 & 0 & 0 & 0 & 1 & 0 & 0 & 0\\
\textcolor{red}{1} & 1 & 1 & 0 & 0 & 1 & 1 & 1 & 0 & 0\\
\textcolor{red}{1} & 1 & 0 & 1 & 1 & 1 & 1 & 0 & 1 & 0\\
\textcolor{red}{1} & 0 & 0 & 0 & 0 & 0 & 0 & 0 & 1 & 1\\
  \textcolor{red}{0} & 1 & 0 & 0 & 0 & 0 & 0 & 1 & 0 & 1\\
\textcolor{red}{1} & 1 & 1 & 0 & 0 & 0 & 1 & 1 & 0 & 0\\
\textcolor{red}{1} & 1 & 0 & 1 & 0 & 1 & 0 & 0 & 1 & 0\\
\textcolor{red}{1} & 0 & 0 & 1 & 0 & 0 & 1 & 1 & 1 & 1\\
  \textcolor{red}{0} & 1 & 1 & 1 & 1 & 1 & 0 & 1 & 1 & 1\\
\textcolor{red}{1} & 0 & 1 & 1 & 0 & 1 & 0 & 0 & 1 & 1\\
  \textcolor{red}{0} & 0 & 0 & 0 & 0 & 0 & 1 & 1 & 0 & 1\\
  \textcolor{red}{0} & 0 & 0 & 0 & 0 & 1 & 0 & 0 & 0 & 0\\
  \textcolor{red}{0} & 0 & 0 & 0 & 1 & 0 & 1 & 0 & 0 & 0\\
  \textcolor{red}{0} & 0 & 0 & 1 & 0 & 0 & 1 & 1 & 0 & 0\\
  \textcolor{red}{0} & 0 & 1 & 1 & 1 & 1 & 0 & 0 & 1 & 0\\
  \textcolor{red}{0} & 1 & 0 & 1 & 0 & 1 & 1 & 1 & 1 & 1\\
\textcolor{red}{1} & 1 & 0 & 1 & 0 & 1 & 1 & 1 & 1 & 1\\
\textcolor{red}{1} & 0 & 0 & 1 & 0 & 1 & 1 & 1 & 1 & 1\\
  \textcolor{red}{0} & 1 & 1 & 1 & 0 & 1 & 1 & 1 & 1 & 1\\
\textcolor{red}{1} & 0 & 1 & 0 & 0 & 1 & 1 & 1 & 1 & 1\\
  \textcolor{red}{0} & 0 & 1 & 1 & 1 & 1 & 1 & 1 & 1 & 1\\
  \textcolor{red}{0} & 1 & 0 & 1 & 0 & 0 & 1 & 1 & 1 & 1\\
\textcolor{red}{1} & 1 & 0 & 1 & 1 & 1 & 0 & 1 & 1 & 1\\
\textcolor{red}{1} & 0 & 0 & 0 & 0 & 1 & 0 & 0 & 1 & 1\\
  \textcolor{red}{0} & 1 & 0 & 0 & 1 & 0 & 1 & 1 & 0 & 1\\
\textcolor{red}{1} & 1 & 1 & 1 & 0 & 0 & 0 & 0 & 0 & 0\\
\textcolor{red}{1} & 1 & 1 & 0 & 1 & 0 & 0 & 0 & 0 & 0\\
\textcolor{red}{1} & 1 & 0 & 0 & 0 & 1 & 0 & 0 & 0 & 0\\
\textcolor{red}{1} & 0 & 1 & 0 & 1 & 0 & 1 & 0 & 0 & 0\\
  \textcolor{red}{0} & 0 & 1 & 0 & 0 & 0 & 1 & 1 & 0 & 0\\
  \textcolor{red}{0} & 1 & 1 & 1 & 0 & 1 & 0 & 0 & 1 & 0\\
\textcolor{red}{1} & 0 & 1 & 0 & 0 & 0 & 1 & 1 & 1 & 1\\
  \textcolor{red}{0} & 0 & 1 & 1 & 0 & 1 & 0 & 1 & 1 & 1\\
  \textcolor{red}{0} & 1 & 0 & 0 & 0 & 0 & 0 & 0 & 1 & 1\\
  \hline
    \end{tabular}
            \end{tiny}}
        \subfloat[\label{tab:CA:52ab }]{
             \begin{tiny}
              \begin{tabular}{|c|c|c|c|c|c|c|c|c|c|}\hline
   150&150&150&150&150&150&150&150&150&150\\\hline
   \textcolor{red}{1} & 0 & 1 & \pmb{1} & 1 & 1 & 1 & 1 & 1 & 0\\
   \textcolor{red}{1} & 0 & 0 &  \pmb{1} & 1 & 1 & 1 & 1 & 0 & 1\\
   \textcolor{red}{1} & 1 & 1 & \pmb{0} & 1 & 1 & 1 & 0 & 0 & 1\\
    \textcolor{red}{0} & 1 & 0 & \pmb{0} & 0 & 1 & 0 & 1 & 1 & 1\\
   \textcolor{red}{1} & 1 & 1 & \pmb{0} & 1 & 1 & 0 & 0 & 1 & 0\\
    \textcolor{red}{0} & 1 & 0 & \pmb{0} & 0 & 0 & 1 & 1 & 1 & 1\\
   \textcolor{red}{1} & 1 & 1 & \pmb{0} & 0 & 1 & 0 & 1 & 1 & 0\\
    \textcolor{red}{0} & 1 & 0 &  \pmb{1} & 1 & 1 & 0 & 0 & 0 & 1\\
   \textcolor{red}{1} & 1 & 0 & \pmb{0} & 1 & 0 & 1 & 0 & 1 & 1\\
    \textcolor{red}{0} & 0 & 1 &  \pmb{1} & 1 & 0 & 1 & 0 & 0 & 0\\
    \textcolor{red}{0} & 1 & 0 &  \pmb{1} & 0 & 0 & 1 & 1 & 0 & 0\\
   \textcolor{red}{1} & 1 & 0 &  \pmb{1} & 1 & 1 & 0 & 0 & 1 & 0\\
    \textcolor{red}{0} & 0 & 0 & \pmb{0} & 1 & 0 & 1 & 1 & 1 & 1\\
    \textcolor{red}{0} & 0 & 0 &  \pmb{1} & 1 & 0 & 0 & 1 & 1 & 0\\
    \textcolor{red}{0} & 0 & 1 & \pmb{0} & 0 & 1 & 1 & 0 & 0 & 1\\
    \textcolor{red}{0} & 1 & 1 &  \pmb{1} & 1 & 0 & 0 & 1 & 1 & 1\\
   \textcolor{red}{1} & 0 & 1 &  \pmb{1} & 0 & 1 & 1 & 0 & 1 & 0\\
   \textcolor{red}{1} & 0 & 0 & \pmb{0} & 0 & 0 & 0 & 0 & 1 & 1\\
   \textcolor{red}{1} & 1 & 0 & \pmb{0} & 0 & 0 & 0 & 1 & 0 & 0\\
   \textcolor{red}{0} & 0 & 1 & \pmb{0} & 0 & 0 & 1 & 1 & 1 & 0\\
    \textcolor{red}{0} & 1 & 1 &  \pmb{1} & 0 & 1 & 0 & 1 & 0 & 1\\
   \textcolor{red}{1} & 0 & 1 & \pmb{0} & 0 & 1 & 0 & 1 & 0 & 1\\
   \textcolor{red}{1} & 0 & 1 &  \pmb{1} & 1 & 1 & 0 & 1 & 0 & 1\\
   \textcolor{red}{1} & 0 & 0 &  \pmb{1} & 1 & 0 & 0 & 1 & 0 & 1\\
   \textcolor{red}{1} & 1 & 1 & \pmb{0} & 0 & 1 & 1 & 1 & 0 & 1\\
    \textcolor{red}{0} & 1 & 0 &  \pmb{1} & 1 & 0 & 1 & 0 & 0 & 1\\
   \textcolor{red}{1} & 1 & 0 & \pmb{0} & 0 & 0 & 1 & 1 & 1 & 1\\
    \textcolor{red}{0} & 0 & 1 & \pmb{0} & 0 & 1 & 0 & 1 & 1 & 0\\
    \textcolor{red}{0} & 1 & 1 &  \pmb{1} & 1 & 1 & 0 & 0 & 0 & 1\\
   \textcolor{red}{1} & 0 & 1 &  \pmb{1} & 1 & 0 & 1 & 0 & 1 & 1\\
   \textcolor{red}{1} & 0 & 0 &  \pmb{1} & 0 & 0 & 1 & 0 & 0 & 0\\
   \textcolor{red}{1} & 1 & 1 &  \pmb{1} & 1 & 1 & 1 & 1 & 0 & 0\\
    \textcolor{red}{0} & 1 & 1 &  \pmb{1} & 1 & 1 & 1 & 0 & 1 & 0\\
   \textcolor{red}{1} & 0 & 1 &  \pmb{1} & 1 & 1 & 0 & 0 & 1 & 1\\
   \textcolor{red}{1} & 0 & 0 &  \pmb{1} & 1 & 0 & 1 & 1 & 0 & 0\\
   \textcolor{red}{1} & 1 & 1 & \pmb{0}& 0 & 0 & 0 & 0 & 1 & 0\\
    \textcolor{red}{0} & 1 & 0 &  \pmb{1} & 0 & 0 & 0 & 1 & 1 & 1\\
   \textcolor{red}{1} & 1 & 0 &  \pmb{1} & 1 & 0 & 1 & 0 & 1 & 0\\
    \textcolor{red}{0} & 0 & 0 & \pmb{0} & 0 & 0 & 1 & 0 & 1 & 1\\
    \textcolor{red}{0} & 0 & 0 & \pmb{0} & 0 & 1 & 1 & 0 & 0 & 0\\
    \textcolor{red}{0} & 0 & 0 & \pmb{0} & 1 & 0 & 0 & 1 & 0 & 0\\
   \textcolor{red}{0} & 0 & 0 &  \pmb{1} & 1 & 1 & 1 & 1 & 1 & 0\\
    \textcolor{red}{0} & 0 & 1 & \pmb{0} & 1 & 1 & 1 & 1 & 0 & 1\\
    \textcolor{red}{0} & 1 & 1 & \pmb{0} & 0 & 1 & 1 & 0 & 0 & 1\\
   \textcolor{red}{1} & 0 & 0 &  \pmb{1} & 1 & 0 & 0 & 1 & 1 & 1\\
   \textcolor{red}{1} & 1 & 1 & \pmb{0} & 0 & 1 & 1 & 0 & 1 & 0\\
    \textcolor{red}{0} & 1 & 0 &  \pmb{1} & 1 & 0 & 0 & 0 & 1 & 1\\
   \textcolor{red}{1} & 1 & 0 & \pmb{0} & 0 & 1 & 0 & 1 & 0 & 0\\
    \textcolor{red}{0} & 0 & 1 & \pmb{0} & 1 & 1 & 0 & 1 & 1 & 0\\
    \textcolor{red}{0} & 1 & 1 & \pmb{0} & 0 & 0 & 0 & 0 & 0 & 1\\
   \textcolor{red}{1} & 0 & 0 &  \pmb{1} & 0 & 0 & 0 & 0 & 1 & 1\\
  \textcolor{red}{1} & 1 & 1 &  \pmb{1} & 1 & 0 & 0 & 1 & 0 & 0\\
    \textcolor{red}{0} & 1 & 1 &  \pmb{1} & 0 & 1 & 1 & 1 & 1 & 0\\
   \textcolor{red}{1} & 0 & 1 & \pmb{0} & 0 & 0 & 1 & 1 & 0 & 1\\
   \textcolor{red}{1} & 0 & 1 &  \pmb{1} & 0 & 1 & 0 & 0 & 0 & 1\\
   \textcolor{red}{1} & 0 & 0 & \pmb{0} & 0 & 1 & 1 & 0 & 1 & 1\\
   \textcolor{red}{1} & 1 & 0 & \pmb{0} & 1 & 0 & 0 & 0 & 0 & 0\\
    \textcolor{red}{0} & 0 & 1 &  \pmb{1} & 1 & 1 & 0 & 0 & 0 & 0\\
    \textcolor{red}{0} & 1 & 0 &  \pmb{1} & 1 & 0 & 1 & 0 & 0 & 0\\
   \textcolor{red}{1} & 1 & 0 & \pmb{0} & 0 & 0 & 1 & 1 & 0 & 0\\
    \textcolor{red}{0} & 0 & 1 & \pmb{0} & 0 & 1 & 0 & 0 & 1 & 0\\
    \textcolor{red}{0} & 1 & 1 &  \pmb{1} & 1 & 1 & 1 & 1 & 1 & 1\\
  \hline
    \end{tabular}
            \end{tiny} }
\end{table}

 As was the case with the 102-CA, the vertical sequences generated by these 150/90-CA are also interleaving sequences.
 
\begin{theorem}
All   sequences in the 150/90-CA are  interleaving sequences composed of shifted versions of $\{a_i\}$ or/and the zero column.
 \end{theorem}
\begin{proof}
    We consider the case of 2-interleaving sequences. The general case of  $2^t$-interleaving is just a generalisation of the arguments below.

    Consider the 2-interleaving sequence given by:
$$\{a_0,a_k,a_1,a_{k+1}, a_2, a_{k+2} \ldots \} 
   $$
   Assume this sequence appears in the 0-th column of the CA.
   If rule 90 controls this column, then the next sequence is the same sequence just shifted one position (see Table~\ref{tab:150902intera}).
   If rule 150 controls this column, then the next sequence has the form
   $$\{a_0+a_k,a_1+a_k,a_1+a_{k+1}, a_2+a_{k+1},\ldots\}$$
   (see Table~\ref{tab:150902interb}), that is, a  2-interleaving sequence of the sequences $\{a_i+a_{i+k}\}$, $\{a_{i+1}+a_{i+k}\}$,
   which we know are shifted versions of 
   the PN-sequence $\{a_i\}$ (see Corollary~\ref{cor:2}) or the null sequence. 

   Now, assume we have two 2-interleaving sequences
   \begin{align*}
       \{a_0,a_k,a_1,a_{k+1},a_2,a_{k+1},\ldots\}\\
       \{a_{k_1},a_{k_2},a_{k_1+1},a_{k_2+1},a_{k_1+2},a_{k_2+1}\ldots \}\\
   \end{align*}
   in the middle of the CA. 
   If rule 90 controls this column, then the next sequence has the following form
   $$\{a_0+a_{k_2}, a_k+a_{k_1+1}, a_1+a_{k_2+1}, a_{k+1}+a_{k_1+2},a_2+a_{k_2+2}, a_{k+2}+a_{k_1+3},\ldots\}$$
    (see Table~\ref{tab:150902interc}).
  This sequence is a 2-interleaving sequence composed of the sequences
  $\{a_i+a_{i+k_2}\}$, $\{a_{i+k}+a_{i+1+k_1}\}$,  which   are shifted versions of 
   the PN-sequence $\{a_i\}$ (Corollary~\ref{cor:2}) or the null sequence.

   If rule 150 controls this column, then the next sequence has the following form
   $$\{a_0+a_{k_1}+a_{k_2}, a_k+a_{k_1+1}+a_{k_2}, a_1+a_{k_1+1}+a_{k_2+1}, a_{k+1}+a_{k_1+2}+a_{k_2+1}, a_2+a_{k_1+2}+a_{k_2+2}, a_{k+2}+a_{k_1+3}+a_{k_2+2} \ldots\}$$
    (see Table~\ref{tab:150902interd}).
  This sequence is a 2-interleaving sequence composed of the sequences
  $\{a_i+a_{i+k_1}+a_{i+k_2}\}$, $\{a_{i+k}+a_{i+1+k_1}+a_{i+k_2}\}$,  which   are shifted versions of 
   the PN-sequence $\{a_i\}$ (Corollary~\ref{cor:2}) or the null sequence. 

   As a consequence, we claim that all sequences in the CA are 2-interleaving sequences composed by interleaving shifted versions of the same PN-sequence and sometimes the zero sequence.    
\end{proof}

\begin{table}
    \centering
    \subfloat[\label{tab:150902intera}]
    {
  $ \begin{array}{|c|c|}\hline
    90&-\\\hline\hline
      a_0  &  a_k\\
      a_k  & a_1\\
      a_1  & a_{k+1}\\
   a_{k+1} & a_2\\
      \vdots &\vdots\\\hline
    \end{array}$
    }
    \quad
    \subfloat[\label{tab:150902interb}]{
    $\begin{array}{|c|c|} \hline
        150&-\\\hline\hline
      a_0   &  a_0+a_k\\
      a_k   & a_1+a_k\\
      a_1 & a_1+a_{k+1}\\
      a_{k+1}&  a_2+a_{k+1}\\
      \vdots &\vdots\\\hline
    \end{array}$}

    \subfloat[\label{tab:150902interc}]{
    $\begin{array}{|c|c|c|} \hline
        -&90&-\\\hline\hline
      a_{0}    &   a_{k_1} &     a_{0} +a_{k_2}   \\
      a_{k}  & a_{k_2} &   a_{k} +a_{k_1+1} \\
      a_{1}  & a_{k_1+1} &   a_{1} +a_{k_2+1} \\
      a_{k+1} & a_{k_2+1}&a_{k+1}+a_{k_1+2} \\
      \vdots &\vdots&\vdots\\\hline
    \end{array}$}
    \quad
    \subfloat[\label{tab:150902interd}]{
    $\begin{array}{|c|c|c|} \hline
        -&150&-\\\hline\hline
      a_{0}    &   a_{k_1} &     a_{0} +a_{k_1} +a_{k_2}   \\
      a_{k}  & a_{k_2} &   a_{k} +a_{k_2}+a_{k_1+1}  \\
      a_{1}  & a_{k_1+1} &   a_{1} +a_{k_1+1} +a_{k_2+1} \\
      a_{k+1} & a_{k_2+1}&a_{k+1}+a_{k_2+1} ++a_{k_1+2} \\
      \vdots &\vdots&\vdots\\\hline
    \end{array}$}
   
    \caption{Behaviour of rules 90 and 150 in the CA}
    \label{tab:150902inter}
\end{table}

\begin{remark}
     In both cases, when an interleaving sequence in the CA is created by interleaving a sequence of zeros with a PN-sequence, the linear complexity and period remain consistent with the given 2-interleaving sequence. However, the number of zeros in the sequence is noticeably higher than the number of ones, making it less favourable as a potential keystream compared to the others.  
\end{remark}
 
  \begin{example}
      Consider the second
CA in Table~\ref{tab:CA:52a}. 
 The 2-interleaving sequence given in Example~\ref{ex:5:2:1} is represented in the 0-th column. Consider now another sequence of this CA, for example, the one in the third column (in bold):
 $$\{  1   1   0   0   0   0   0   1   0   1   1   1   0   1   0   1   1   0   0   0   1   0   1   1   0   1   0   0   1   1   1   1   1   1   1   0   1   1   0   0 0   1   0   0   1   0   1   0   0   0   1   1   1   0   1   0   0   1   1   0   0   1\}$$
 This sequence is composed of the PN-sequences
\begin{align*}
  \{   1   0   0   0   0   1   0   0   1   0   1   1   0   0   1   1   1   1   1   0   0   0   1   1   0   1   1   1   0   1   0\},\\
  \{   1   0   0   1   1   1   1   1   0   0   0   1   1   0   1   1   1   0   1   0   1   0   0   0   0   1   0   0   1   0   1\},
\end{align*}
therefore is a 2-interleaving sequence 
of shifted versions of the PN-sequence considered in Example~\ref{ex:5:2:1}.  \end{example}

Similar to the case of the 102-CAs, exploring whether 150/90-CAs generate interleaving sequences when $t$ is not a power of two would be intriguing. This particular scenario seems intricate and is best left for future research work. Additionally, there is currently a lack of evidence regarding the existence of such CAs, indicating the necessity for a more thorough investigation.
\section{Comparison}\label{sec:comp}

 Comparing both families  of 102-CAs and 150/90-CAs, the study reveals notable distinctions in their characteristics. The 102-CAs consistently maintain a uniform structure, making them straightforward to assemble due to their inherent regularity. In contrast, the 150/90-CAs pose  a more intricate challenge, necessitating the implementation of the Cattell-Muzio algorithm for construction. 
Moreover, the length of the 102-CAs is significantly greater than that for the 150/90-CAs, showing an exponential difference in size. 
Indeed, when working with $2^t$-interleaving sequences  generated by interleaving PN-sequences of period $T=2^L-1$, the length of both   150/90-CAs is typically $t\cdot L$, which is linear in $L$. In the case of the 102-CA, the length is given by $\frac{2^L-1}{\gcd(2^L-1,D)}$, which is usually exponential in $L$.

Despite their differences, both families share common ground in generating vertical sequences with identical properties. The consistency in the characteristics of these vertical sequences establishes a notable similarity between the two cellular automata models, despite their divergent construction complexities and   lengths.

 \section{Conclusions}\label{sec:concl} 

This paper introduces two families of CA that stand out for their ability to produce interleaving sequences. We conducted an analysis of the structure and length of both families, along with the sequences they generate.
In addition, we conducted a  comparison of the two families, highlighting their key similarities and differences. It is worth noting that the considered interleaving sequences in this study are composed of translated versions of the same PN-sequence. It is important to note that previous studies have identified these sequences as strong candidates for cryptographic applications. This approach provides a deeper insight into the nature and properties of the sequences generated by these CA.
In summary, our investigation reaffirms the well-established fact that Cellular Automata   possess the ability to generate cryptographic sequences.

\section*{Acknowledgements}
This work was supported by CNPq Brazil with   process 405842/2023-6  and by FAPESP with process 2024/05051-7.
 
\bibliographystyle{elsarticle-num}
\bibliography{biblio}


\section*{Appendix 1}
\begin{landscape}
  \begin{table}
\caption{102-CA that generates a 2-interleaving sequence  \label{tab:gen:102CAa}}
\centering
\[
\begin{array}{|c|c|c|c|c|c|c|c|}\hline
\textbf{102} & \textbf{102} & \textbf{102} & \textbf{102} & \textbf{102} & \textbf{102}  & \textbf{102} & \ldots \\\hline\hline
a_0 & a_0+a_k & a_0+a_1 & a_0+a_1+a_k+a_{k+1} & a_0+a_2&a_0+a_2+a_k+a_{k+2}  &  a_0+a_1+a_2+a_3  & \ldots\\
 a_k & a_k+a_{1} & a_k+a_{k+1} & a_k+a_{k+1}+a_1+a_2 & a_k+a_{k+2} & a_k+a_{k+2} +a_1+a_3&    a_k+a_{k+1}+ a_{k+2}+a_{k+3}& \ldots\\
 a_1 & a_{1}+a_{k+1}&a_1+a_2&a_1+a_2+a_{k+1}+a_{k+2}&a_1+a_3&a_1+a_3+a_{k+1}+a_{k+3}&a_1+a_2+a_3+a_4& \ldots\\
 a_{k+1} & a_{k+1} +a_{2}& a_{k+1}+a_{k+2}&a_{k+1}+a_{k+2}+a_2+a_3 &a_{k+1}+a_{k+3}&a_{k+1}+a_{k+3}+a_2+a_4&a_{k+1}+ a_{k+2}+a_{k+3}+a_{k+4} &\ldots\\
\vdots & \vdots & \vdots & \vdots & \vdots &  & \vdots &   \\\hline
\end{array}
\]
\end{table} 
\end{landscape}
\begin{landscape}
  \begin{table}
\caption{CA that generates a 4-interleaving sequence  with shifted versions of the same PN-sequence\label{tab:gen:102CAb}}
\centering
\[
\begin{array}{|c|c|c|c|c|c|}\hline
\textbf{102} & \textbf{102}& \textbf{102}    & \textbf{102} & \textbf{102}  & \ldots \\\hline\hline
 a_0     & a_0+a_{k_1}     & a_0+a_{k_2}     & a_0+a_{k_1}+a_{k_2}+a_{k_3}    & a_0+a_1&          \ldots\\
 a_{k_1} & a_{k_1}+a_{k_2} & a_{k_1}+a_{k_3} &  a_{k_1}+a_{k_2}+a_{k_3}+a_1   &  a_{k_1}+ a_{k_1+1}   &\ldots\\
 a_{k_2} & a_{k_2}+a_{k_3} &a_{k_2}+a_1      & a_{k_2}+a_{k_3}+a_1 +a_{k_1+1} & a_{k_2}+ a_{k_2+1}   & \ldots\\
 a_{k_3} & a_{k_3} +a_{1}  &a_{k_3}+a_{k_1+1}& a_{k_3}+a_1+a_{k_1+1}+a_{k_2+1}& a_{k_3}+ a_{k_2+1}    & \ldots\\
 a_1     & a_1+a_{k_1+1}   &a_1+a_{k_2+1}    & a_1+a_{k_1+1}+a_{k_2+1}+a_{k_3+1} &  a_1+a_2                   & \ldots\\
 a_{k_1+1}&a_{k_1+1}+a_{k_2+1}&a_{k_1+1}+a_{k_3+1}&  a_{k_1+1}+a_{k_2+1}+a_{k_3+1}+a_2  &    a_{k_1+1}+ a_{k_1+2}   & \ldots\\
 a_{k_2+1}&a_{k_2+1}+a_{k_3+1}&a_{k_2+1}+a_2      & a_{k_2+1}+a_{k_3+1}+a_2 +a_{k_1+2}  & a_{k_2+1}+ a_{k_2+2}                  &  \ldots\\
 a_{k_3+1}&a_{k_3+1}+a_{2}    &a_{k_3+1}+a_{k_1+2}      & a_{k_3+1}+a_2+a_{k_1+2}+a_{k_2+2}  &a_{k_3+1}+ a_{k_2+2} & \ldots\\
\vdots & \vdots & \vdots & \vdots & \vdots &    \\\hline
\end{array}
\]
\end{table}

\begin{table}
\caption{CA that generates a 2-interleaving sequence fo two different PN-sequences \label{tab:gen:102CA}}
\centering
\[
\begin{array}{|c|c|c|c|c|c|c|c|}\hline
\textbf{102} & \textbf{102} & \textbf{102} & \textbf{102} & \textbf{102} & \textbf{102}  & \textbf{102} & \ldots \\\hline\hline
a_0 & a_0+b_0 & a_0+a_1 & a_0+a_1+b_0+b_1 & a_0+a_1+a_2+a_3 &a_0+a_1+a_2+a_3+  b_0+b_1+b_2+b_3  &  a_0+a_4  & \ldots\\
 b_0 & b_0+a_1 & b_0+b_1 & b_0+b_1+a_1+a_2 & b_0+b_1+b_2+b_3 &  b_0+b_1+b_2+b_3+ a_1+a_2+a_3+a_4& b_0+b_4   & \ldots\\
a_1 & a_1+b_1 & a_1+a_2 & a_1+a_2+b_1+b_2 & a_1+a_2+a_3+a_4 &a_1+a_2+a_3+a_4+  b_2+b_2+b_3+b_4  &  a_2+a_5  & \ldots\\
b_1 & b_1+a_2 & b_1+b_2 & b_1+b_2+a_2+a_3 & b_1+b_2+b_3+b_4 &  b_1+b_2+b_3+b_4+ a_2+a_3+a_4+a_5& b_1+b_5   & \ldots\\
\vdots & \vdots & \vdots & \vdots & \vdots &  & \vdots &   \\\hline
\end{array}
\]
\end{table}
\end{landscape}

\end{document}